\documentclass[aps,prl,secnumarabic,twocolumn,
groupedaddress,showpacs,amsmath,amssymb]{revtex4}
\usepackage[cp1251]{inputenc}

\usepackage{bm}
\usepackage{graphicx}

\begin{document}

\title{Formation of a condensed state with macroscopic number of phonons
in ultracold Bose gases}
\author{Yu. Kagan and L.A. Manakova}
\email[]{manakova@kurm.polyn.kiae.su} \affiliation{ RRC "Kurchatov
Institute", 123182 Kurchatov Sq.,1, Moscow, Russia}

\begin{abstract}
A mechanism for the formation of a new type of stationary state
with macroscopical number of phonons in condensed atomic gases is
proposed. This mechanism is based on generating longitudinal
phonons as a result of parametric resonance caused by a permanent
modulation of the transverse trap frequency in an elongated trap.
The phonon-phonon interaction predetermines the self-consistent
evolution which is completed with macroscopic population of one
from all levels within the energy interval of parametric
amplification. This level proves to be shifted to the edge of this
interval. All other levels end the evolution with zero population.

\end{abstract}
\pacs{03.75.Kk, 03.75.Nt, 67.40.Db}

\maketitle

\section {I. Introduction}

At zero temperature the formation of the stationary state in an
atomic condensate with involving of macroscopical number of
phonons certainly requires thermodynamically non-equilibrium
conditions.

In the present paper  an idea is proposed for realizing these
conditions for longitudinal sound phonons in elongated
Bose-condensed ultracold gases. The idea is based on generating
phonons as a result of the parametric resonance induced by a
permanent modulation of the transverse trap frequency
$\omega_{\bot}$. In last years, it has become clear that the
parametric resonance plays an important role in a number of
phenomena in ultracold gases. As is shown in \cite{K1}, the
parametric resonance, responsible for the effective energy
transfer between different branches of excitations, determines the
temporal evolution of the system, in particular, origin of the
damping for the transverse breathing mode at zero temperature. The
further consideration of the nonlinear evolution accompanying the
parametric resonance \cite{K2} allows to explain the nontrivial
picture of the damping revealed in the remarkable work of the
Paris group \cite{Ch}. The parametric amplification of  the
Bogoliubov modes due to periodic modulation of an optical 1D
lattice is considered in \cite{D1}. The interesting picture of the
density distribution in low-dimensional \cite{St} and toroidal
\cite{D2} Bose-condensed gases under conditions of the parametric
excitation has been revealed at the numerical solution of the
Gross-Pitaevskii equation.

A permanent modulation of the transverse trap frequency
$\omega_{\bot}$ brings about the oscillations of the condensate
density and, accordingly, sound velocity. Owing to the parametric
resonance, this results in the generation of pairs of sound
phonons with the opposite momenta and energies close to a half of
the modulation frequency $\omega_0/2$. In fact, phonon pairs are
produced within the finite energy interval $\omega_0/2\pm E_0$. In
what follows, this interval of parametric amplification with the
width $2E_0$ near $\omega_0/2$ is identified as "the parametric
interval". The parameter $E_0$ is connected with the modulation
amplitude $|\delta\omega_{\bot}/\omega_{\bot}|\ll 1$ by the
relation $E_0\sim
\omega_0\cdot|\delta\omega_{\bot}/\omega_{\bot}|$. If $E_0$
exceeds the damping factor $\gamma$ for longitudinal phonons,
their occupation numbers begin to increase exponentially. In this
case the phonon-phonon interaction becomes significant. As a
result, the evolution scenario changes drastically.

For the typical longitudinal size of traps $L$ a few phonon levels
fall into the narrow energy interval $2E_0$. The main effect of
the phonon-phonon interaction, in addition to the damping,
manifests itself in an effective renormalization of the levels. In
the course of the self-consistent nonlinear evolution the position
of levels changes continuously  with changing the number of
phonons. As is shown below, in the simplest case of one
doubly-degenerate level in the parametric interval $2E_0$, the
evolution is over as the renormalized level reaches the left or
right boundary of the interval. Then the growth of the phonon
occupation numbers ceases. As a result, the steady state of
phonons with the macroscopic occupation number in a single state
and the common phase for all excitations appears.

In the case of several levels the picture is more sophisticated.
It turns out that only one of the phonon levels finishes the
evolution achieving the stationary state with the macroscopic
number of phonons. As in the case of a single level, this level
proves to be at the edge of the parametric interval.

The preliminary analysis of the problem for the case of a single
level within the parametric interval has been represent in our
work [8]. In the present paper we give a comprehensive
consideration of the problem, including the general analysis of
the evolution for arbitrary number of phonon levels within the
parametric interval. There is considered the essential problem of
indrawning outer levels located initially out of the energy
interval of parametric amplification. Since in general case the
evolution is over with the stationary occupation only a single
level, the question about stability of this state is specially
analysed.

\section {II. Parametric generation of phonon pairs}

{\bf 1.} Let us consider the Bose-condensed gas at $T=0$ in a trap
of the cylindrical symmetry with $L\gg R$, where $R$ is the radius
of the condensate. Neglecting the edge effects, we can write the
general equation for the field operator $\hat{\Psi}({\bf r},z,t)$
of atoms in the form
\begin{equation} \label{eq:1}
\!i\hbar\frac{\partial \hat{\Psi}}{\partial
t}\!=\!\!\Big[\!\!-\frac{\hbar^2}{2m}\nabla^2_{{\bf
r}}\!-\frac{\hbar^2}{2m}\frac{\partial^2}{\partial
z^2}+\!\frac{\omega_{\bot}^2(t){\bf r}^2}{2}\Big]\hat{\Psi}+U_0
\hat{\Psi}^+\hat{\Psi}\hat{\Psi},
\end{equation}
where $U_0=4\pi\hbar^2 a/m$, $a$ being the 3D s-scattering length.
Note, considering a dilute and cold Bose gas with dominant binary
collisions we imply as usually that the radius of interaction
region $r_0\ll a$, the gas parameter $na^3\ll 1$, and the
correlation length $\xi> n^{-1/3}$ ($n$ is the atomic density). In
this case, the replacement of two-body interaction with the
effective contact potential is relevant. We assume that the trap
transverse frequency depends on time as
\begin{equation} \label{eq:2}
\omega_{\bot}(t)=\omega_{\bot}(1+\eta\sin\omega_0 t),\;\;\;\eta\ll
1.
\end{equation}
According to Ref. \cite{K3}, we introduce the spatial scaling
parameter $b(t)$ and new variables $\overrightarrow{\rho}={\bf
r}/b(t)$, $\tau(t)$. The field operator in terms of new variables
can be written as
\begin{equation} \label{eq:3}
\hat{\Psi}=\frac{\hat{\chi}(\overrightarrow{\rho},z,\tau)}{b(t)}\cdot
e^{i\Phi};\;\;\;\;\Phi=\frac{mr^2}{2\hbar}\cdot\frac{\dot{b}}{b};
\end{equation}
Substituting these expressions into Eq.(\ref{eq:1}), we obtain
straightforwardly
\begin{equation} \label{eq:4}
\!\!i\hbar\frac{\partial \hat{\chi}}{\partial
\tau}\!=\!\!\Big[\!\!-\frac{\hbar^2}{2m}\nabla^2_{{\bf
\rho}}\!+\!\frac{m\omega^2_{\bot}\bf{\rho}^2}{2}\Big]\hat{\chi}\!+U_0
\hat{\chi}^+\hat{\chi}\hat{\chi}\!-b^2(t)\frac{\hbar^2
}{2m}\frac{\partial^2\hat{\chi}}{\partial z^2},
\end{equation}
provided that functions $b(t)$ and $\tau(t)$ satisfy the equations
\begin{equation} \label{eq:5}
\frac{d^2
b}{dt^2}+\omega_{\bot}^2(t)b=\frac{\omega^2_{\bot}}{b^3};\;\;\;\;
b^2\frac{d\tau}{dt}=1.
\end{equation}
Treating the condensate wave function as independent of $z$, we
find that Eq.(\ref{eq:4}) being expressed in variables  ${\bf
\rho}, \tau$ describes the evolution at the time-independent
frequency $\omega_{\bot}$. Combining the expression (\ref{eq:2})
with the condition $\omega_0\ll\omega_{\bot}$, we find the
solution of Eq. (\ref{eq:5}) in the form
\begin{equation} \label{eq:6}
b(t)=1+b_1(t),\;\;\;\;\; b_1(t)\approx \frac{-\eta}{2}\cdot
\sin\omega_0 t.
\end{equation}
The field operator $\hat{\chi}$ can be represented in the usual
form $\hat{\chi}=(\chi_0+\hat{\chi}')\exp(-i\mu\tau)$, $\chi_0$
being the condensate wave function and $\mu$ is the initial
chemical potential. In the absence of excitations at $T=0$ we can
omit the last term in Eq. (\ref{eq:4}) and  determine
$\chi_0(\rho)$ from the equation
\begin{equation} \label{eq:4a}
\Big[-\frac{\hbar^2}{2m}\nabla^2_{{\bf
\rho}}+\frac{m\omega^2_{\bot}\bf{\rho}^2}{2}-\mu\Big]\chi_0+U_0
\chi_0^3=0,
\end{equation}
where $\mu$ is the chemical potential corresponding to the static
cylindrical trap with $\omega_{\bot}=const$. In the Thomas-Fermi
limit when $nU_0\gg \hbar\omega_{\bot}$ the solution of this
equation is reduced to known expression
$$\chi_0^2=\frac{\mu}{U_0}\Big(1-\frac{r^2}{R_{\bot}^2b^2(t)}\Big);\;\;\;R_{\bot}^2=\frac{2\mu}{m\omega_{\bot}^2}.$$
In the quasi-1D case when $nU_0\ll \hbar\omega_{\bot}$, we have
$$\chi_0^2=\frac{n_1}{\pi
l_{\bot}^2}\exp\Big(-\frac{r^2}{l_{\bot}^2b^2(t)}\Big);\;\;\;l_{\bot}^2=\frac{\hbar}{m\omega_{\bot}}.$$
In the both cases the modulation of the trap transverse frequency
results in the oscillations of the condensate density.

{\bf 2.} Considering excited states, we first linearize Eq.
(\ref{eq:4}) in $\hat{\chi}'$. Then from Eq. (4) we have
\begin{equation} \label{eq:4b}
\begin{split}
&i\hbar\frac{\partial \hat{\chi}'}{\partial
\tau}=h_0\hat{\chi}'+G(\hat{\chi}'+\hat{\chi}^{'+})-b^2(t)\frac{\hbar^2
}{2m}\frac{\partial^2\hat{\chi}'}{\partial
z^2},\\&h_0=-\frac{\hbar^2}{2m}\nabla^2_{{\bf
\rho}}+\frac{m\omega^2_{\bot}\bf{\rho}^2}{2}+G-\mu;\;\;\;G=U_0
\chi_0^2(\overrightarrow{\rho}).
\end{split}
\end{equation}

Under condition $\omega_0\ll\omega_{\bot}$ only long-wave
longitudinal phonons prove to be involved into the evolution of
the system.

For these phonons the condition $kR\ll 1$ is fulfilled and the
transverse distribution is close to $\chi_0$ (see, e.g.,
\cite{Z}). Comparison of Eqs. (7) and (8) allows us to simplify
the latter equation omitting the first term in the right side.
Using the second relation in (\ref{eq:5}) and going over to
variable $t$, we arrive at the following equation
\begin{equation} \label{eq:10a}
i\hbar\frac{\partial \hat{\chi}'}{\partial
t}=\frac{G}{b^2(t)}(\hat{\chi}'+\hat{\chi}^{'+})-\frac{\hbar^2
}{2m}\frac{\partial^2\hat{\chi}'}{\partial z^2}.
\end{equation}
Involving the symmetry of the problem, we have for the operator
$\hat{\chi}'$ in second quantization
\begin{equation} \label{eq:11}
\hat{\chi}'=\sum\limits_k \chi_k
\hat{a}_k;\;\;\;\;\chi_k=\frac{e^{ikz}}{\sqrt{L}}\cdot\phi_0(\overrightarrow{
\rho}),
\end{equation}
where $\hat{a}_k$ being the annihilation operator for atoms. At
$\omega_0\ll\omega_{\bot}$ the function $\phi_0$ is actually close
to $\chi_0$ for both the quasi-1D and Thomas-Fermi cases. In the
expression (\ref{eq:11}) we keep the ground state alone for the
transverse motion assuming that higher states are insignificant in
the course of evolution. Substitution of the expression
(\ref{eq:11}) into Eq.(\ref{eq:10a}) gives after the standard
averaging over the radial variables
\begin{equation} \label{eq:12}
\!i\hbar\frac{d\hat{a}_k}{d t}\!=\!\Big(\bar{G}+\frac{\hbar^2
k^2}{2m}\Big)\hat{a}_k+\bar{G}\hat{a}_{-k}^+\!-\eta
\bar{G}\sin\omega_0 t(\hat{a}_k\!+\hat{a}_{-k}^+)
\end{equation}
$\bar{G}=\int
d^2\rho\phi_0^2(\overrightarrow{\rho})G(\overrightarrow{\rho})$.
Let us rewrite this equation in terms of phonon operators using
the usual transformation $\hat{a}_k=u_k \hat{b}_k+v_k
\hat{b}_{-k}^+$. When the last term in Eq. (\ref{eq:12}) is
omitted, a set of equations for $\hat{b}_k,\hat{b}_{-k}^+$
determines the well-known Bogoliubov spectrum $\omega_k$ (with the
replacement $ nU_0\rightarrow \bar{G}$). Considering the
nonstationary case, we introduce the substitution
$\hat{b}_k=\hat{\tilde{b}}_k \exp(-i\omega_0 t/2)$ and take into
account that long times $\omega_0 t\gg 1$ are most interesting for
the analysis. Then, within the resonance approximation, which
implies that we discard the strongly oscillating terms with
$\exp(\pm i\omega_0 t/2)$,  the equations for
$\hat{b}_k,\hat{b}_{-k}^+$ are reduced to the form
\begin{equation} \label{eq:14}
i\frac{d\hat{\tilde{b}}_k}{dt}=\xi_k\hat{\tilde{b}}_k+iE_{0k}
\hat{\tilde{b}}^+_{-k};\;\;-i\frac{d\hat{\tilde{b}}^+_{-k}}{dt}=\xi_k\hat{\tilde{b}}^+_{-k}-iE_{0k}
\hat{\tilde{b}}_{k},
\end{equation}
$E_{0k}=(\eta\bar{G}/2\hbar)[(u_k+v_k)/u_k]$,
$\xi_k=\omega_k-\omega_0/2$. For sound phonons
$|(u_k+v_k)/u_k|\approx\hbar\omega_k/\mu$. Since $E_{0k}\ll
\omega_0$ and we are interested in the energy interval
$|\omega_k-\omega_0/2|\leq 2E_0$, the frequency $\omega_k$ in the
expression for $E_{0k}$ can be replaced by $\omega_0$. In the
other words, $E_{0k}$ is replaced by its value at $k=|k_0|$, where
$|k_0|$ is determined from the condition $\omega_0=2\omega_{k_0}$.
As a result, we have
\begin{equation} \label{eq:15}
E_{0k_0}\approx \eta\omega_0=\omega_0\frac{|\delta
\omega_{\bot}|}{\omega_{\bot}}\equiv E_{0}.
\end{equation}

Substituting the transformation $\hat{\tilde{b}}_k=c_{1k}\exp(
i\varepsilon_{0k} t) \hat{\tilde{b}}^{'}_k+c_{2k}\exp(-
i\varepsilon_{0k} t) \hat{\tilde{b}}_{-k}^{+'}$ into
(\ref{eq:14}), we find $\varepsilon_{0k}=i\alpha_{0k}$ with
$\alpha_{0k}=[|E_{0}|^2-\xi_k^2]^{1/2}$. This result demonstrates
the appearance of the parametric resonance with an exponential
growth of the phonon occupation numbers at $|E_{0}|>|\xi_k|$,
which is induced by the modulation of the transverse trap
frequency. Thus the parametric resonance occurs within the narrow
range near $\omega_0$ with the width $2E_0$.

\section {III. Evolution equations}

{\bf 1.} So far we have disregarded the phonon-phonon interaction.
With an exponential growth of the phonon number the interaction
begins to play an essential role. The weakness of the
phonon-phonon interaction as itself (see below) allows us to take
only three- and four-phonon processes into account. Assuming
generation of sound phonons alone in the parametric interval
$2E_0$, we can use the expressions for $H^{(3)}$ and $H^{(4)}$
obtained within the hydrodynamic approximation, see \cite{L}.
Owing to smallness of the gas parameter, the dominant term in the
Hamiltonian of three-phonon interactions has the form \cite{L}
\begin{equation} \label{eq:16}
H^{(3)}= \frac{m}{2}\int d^3r \hat{{\bf v}}\delta \hat{n}
\hat{{\bf v}},
\end{equation}
Here $\hat{{\bf v}}$, $\delta\hat{n}$ are the operators of
velocity and alternating part of the density, respectively. In
fact, we consider the interaction of longitudinal sound phonons
that effectively reduces $H^{(3)}$ to the one-dimensional problem.

Using the known expressions for operators $\hat{{\bf v}}$, $\delta
\hat{n}$ \cite{L}, one can find the vertex $A_3$ for $H^{(3)}$ in
(14).

Since each three-phonon vertex implies the momentum conservation
law, at least one of three phonons lies beyond the parametric
interval and, therefore, has zero occupation number at $T=0$. This
makes possible to reduce the expression obtained in second order
in $H^{(3)}$ to the effective Hamiltonian for the four-phonon
interaction ($\Delta(k)$ is the Kroneker symbol)
\begin{equation} \label{eq:17}
H_{eff}=\frac{A}{2}\sum\limits_{k_1,...,k_4}
\hat{b}^+_{k_1}\hat{b}^+_{k_2}\hat{b}_{k_3}\hat{b}_{k_4}\Delta(k_1+k_2-k_3-k_4).
\end{equation}
Here all states lie within the interval of $2E_0$. Using the
expression for $A_3$ and taking into account that all wave vectors
$|k_i|\sim k_0$ ($E_0\ll \omega_0$), we find (to the accuracy of
numerical factor of the order of unity)
\begin{equation} \label{eq:II.11a}
|A|\approx\frac{\omega_0^2}{\mu N},
\end{equation}
where $N$ is the total number of particles.

The same four-phonon interaction can be obtained as a result of
the standard canonical transformation of the Hamiltonian
$H_0+H^{(3)}$ (see, for example, \cite{Ag}).

The direct calculation shows that  the ratio of the matrix element
in $H^{(4)}$ from Ref. \cite{L} to $|A|$ is proportional to
$(na^3)^{1/2}\ll 1$.

In addition, second order in $H^{(3)}$ contains the imaginary part
related to real decay processes of phonons, which determine the
phonon damping. The later we take into account phenomenologically
introducing a decrement $\gamma$. After  work \cite{Ch} one can
conclude that the parameter $\gamma$ is small at $T=0$ for the
geometry under consideration. Therefore, it is rather easy to
satisfy the conditions when $\alpha_{0k}>\gamma$ and the
parametric growth of the number of phonons remains.

{\bf 2.} As is follows from the above expressions, the Hamiltonian
of interacting phonons near the parametric resonance have the form
\begin{equation} \label{eq:II.12}
\!\!H\!=\!\sum\limits_k \omega_k \hat{b}_k^+\hat{b}_k+\!\sum_k
\!\!\Big(E_{0}e^{-i\omega_0 t} \hat{b}_k^+
 \hat{b}_{-k}^+ +H.c.\Big) \!+H_{eff}
\end{equation}
Let us find $(d\hat{b}_k/dt)$ taking into account the
phonon-phonon interaction. Using the substitution
$\hat{b}_k=\hat{\tilde{b_k}} \exp(-i\omega_0 t/2)$, we have
\begin{equation} \label{eq:19}
\begin{split}
i\frac{d\hat{\tilde{b}}_k}{dt}=&(-i\gamma_k+\xi_k)\hat{\tilde{b}}_k+iE_0
\hat{\tilde{b}}^+_{-k}+\\ &+
\sum\limits_{k_2k_3k_4}A\hat{\tilde{b}}^+_{k_2}
\hat{\tilde{b}}_{k_3}\hat{\tilde{b}}_{k_4}\Delta(k+k_2-k_3-k_4).
\end{split}
\end{equation}
By means of the equations for $\hat{\tilde{b}}_{k}$ and
$\hat{\tilde{b}}^+_{-k}$ one can directly obtain the equations for
the correlators $N_k=<\hat{\tilde{b}}^+_k \hat{\tilde{b}}_k>$ and
$\hat{f}_k=<\hat{\tilde{b}}_k \hat{\tilde{b}}_{-k}>$. These
equations contain the four-phonon correlators. Within the
mean-field approximation we represent the four-phonon terms as
products of the two-phonon correlators  $N_k$ and $\hat{f}_k$.
This implies that we keep the terms linear in $A$. In addition, we
go over to the classical Bose field for phonons supposing that the
number of parametrically excited phonons is large. As a result, we
arrive at a set of nonlinear equations that describes the
self-consistent evolution of interacting phonons within the
parametric interval. These equations have the form
\begin{equation} \label{eq:20}
\begin{split}
&\frac{dN_k}{dt}\!\!=\!\!\!-2\gamma N_k\!+\!E_{0}
(f_k\!+\!f^*_k)\!+\!iA(\mathcal{P}^{*}\!f_k\!-\!\! \mathcal{P}
f^*_k);\\&\frac{df_k}{dt}\!\!=\!\!-2(\gamma\!+\!i\bar{\xi}_k)f_k\!+\!2E_0
N_k\!-\!2iA\mathcal{P} N_k,
\end{split}
\end{equation}
where the position of the renormalized level is determined by the
expression
\begin{equation} \label{eq:20a}
\bar{\xi}_k=\xi_k+\mathcal{Q}+\bar{A}N_k;\;\;\;\;\bar{A}=3A.
\end{equation}
Here $\xi_k$ is the level position at the initial time moment,
$\mathcal{Q}=2A\sum\limits_{k'\neq \pm k}N_{k'}$;
$\mathcal{P}=\sum\limits_{k'\neq \pm k}f_{k'}$. Hereafter the
evident relations $\omega_k=\omega_{-k}$, $N_k=N_{-k}$ are used.

For the large occupation numbers $N_k$  to the accuracy of terms
$\sim 1/N_k$, the function $f_k(t)$ can be written in the form
$f_k(t)=|f_k(t)|\exp[i\varphi_k(t)]=N_k(t)\exp[i\varphi_k(t)]$. As
a result, we re-arrange (\ref{eq:20}) to a set of equations for
$N_{k}$, $\varphi_{k}$
\begin{equation} \label{eq:C1}
\begin{split}
\!\!\!\!\!\frac{dN_{k}}{dt}=-2\gamma N_{k}+&2E_0
N_{k}\cos\varphi_{k}-\\ & -2AN_{k}\sum\limits_{{k'\neq \pm k}}
N_{k'}\sin(\varphi_{k}-\varphi_{k'});\\
&\!\!\!\!\!\!\!\!\!\!\!\!\!\!\!\!\!\!\!\!\!\!\!\!\!\!\!\!\!\!\!\!\!\!\!\!\!\!\!\!\!\!\!\!\!\!\!\frac{d\varphi_{k}}{dt}\!=\!-2\bar{\xi}_k-2E_0
\sin\varphi_{k}\!\!-2A\!\!\!\sum\limits_{{k'\neq \pm k}}
\!\!N_{k'}\cos(\varphi_{k}-\varphi_{k'}),
\end{split}
\end{equation}
Hereafter we take into consideration that
$\varphi_{k}=\varphi_{-k}$.

In what follows, we consider the case of finite longitudinal size
$L$ assuming that a single or a few discrete two-fold degenerate
levels lie within the energy interval of about $2E_{0}$.

\section{IV. Solution for a single level}

{\bf 1.} In the case of a single level the equations (\ref{eq:C1})
take the form
\begin{equation} \label{eq:21}
\begin{split}
&\frac{dN_{k}}{dt}=-2\gamma N_{k}+2E_0 N_{k}\cos\varphi_{k};\\
&\frac{d\varphi_{k}}{dt}=-2[\xi_k+\bar{A}N_{k}+E_0
\sin\varphi_{k}].
\end{split}
\end{equation}
This system has the following stationary solution
\begin{equation} \label{eq:22}
N_{k}^s=\frac{\sqrt{E_0^2-\gamma^2}\mp\xi_k}{|\bar{A}|};\;\;\;\;
\sin\varphi_{k}^s=\mp\frac{\sqrt{E_0^2-\gamma^2}}{E_0}.
\end{equation}
The signs $\mp$ correspond to $\bar{A}\gtrless0$, respectively.
For definiteness, we suppose $E_0>0$. The result (\ref{eq:22}) has
an interesting physical origin. In fact, the interaction $H_{eff}$
determines the effective renormalization of the phonon level,
leading to $\delta\omega_{k}=\bar{A}N_{k}$ for the case concerned
(see (\ref{eq:20a})). Accordingly, $\xi_k\rightarrow
\bar{\xi}_k=\xi_k+\bar{A}N_{k}$. From Eqs. (\ref{eq:22}) it
straightforwardly follows that
$\bar{\xi}_k^s=\pm\sqrt{E_0^2-\gamma^2}$. So, at $N_k=N_k^s$ the
renormalized level reaches the left or right  edge of the
parametric energy interval (within the accuracy of the shift due
to $\gamma$). As a result, the parametric increase of the phonon
occupation numbers stops and the phase acquires the constant
value. The maximal value of the phonon number equals
\begin{equation} \label{eq:23}
N_{k}^s\approx\frac{E_0}{|\bar{A}|}\gg 1;\;\;\;\;\gamma,\xi_k\ll
E_0.
\end{equation}
The phase is $\varphi_{k}^s\approx\mp\pi/2$ for $\bar{A}\gtrless
0$. The phase $\varphi_k$ corresponds to the phase correlation of
phonon pairs with the opposite momenta. The appearance of the
phase $\varphi_k$ and correlator $f_k$ is connected, first of all,
with the creation of phonon pairs in the course of the evolution
as a result of the parametric resonance. The evolution finishes in
the stationary state of phonon pairs with zero total momentum
and the common phase $\varphi_k^s$.

{\bf 2.} The character of the temporal evolution, tending
asymptotically to the values (\ref{eq:22}), (\ref{eq:23}), depends
essentially on the relations between the parameters in Eqs.
(\ref{eq:21}). First, let $\gamma=0$. In this case, Eqs.
(\ref{eq:21}) have the conserved integral of motion
\begin{equation}\label{eq:24}
H_0=2E_0 N_{k}\sin\varphi_{k}+ \bar{A}N_{k}^2+2\xi_k
N_{k},\;\;\;\;\frac{dH_{0}}{dt}=0.
\end{equation}
The variables $N_{k},\varphi_{k}$ are canonically conjugate, and
the equation of motion (\ref{eq:21}) can be written in the
Hamiltonian form  $$\frac{dN_{k}}{dt}=\frac{\delta
H_{0}}{\delta\varphi_{k}};\;\;\;\;\frac{d\varphi_{k}}{dt}=-\frac{\delta
H_{0}}{\delta N_{k}}.$$At the initial time moment, when the
modulation of $\omega_{\bot}$ is switched on, we suppose
$\bar{\xi}_k(0)=\xi_k+\bar{A}N_{k}(0)$ and
$\bar{\alpha}_{0k}=\sqrt{E_0^2-\bar{\xi}_k(0)^2}$. At that,
Eqs.(\ref{eq:21}) imply that
$\cos\varphi_{k}(0)=\bar{\alpha}_{0k}/E_0$. Bearing in mind that
$\sin\varphi_{k}(0)=-\bar{\xi}_k(0)/E_0$, from the second equation
in (\ref{eq:21}) we obtain $(d\varphi_{k}/dt)(0)=0$. As a
consequence, the initial (and conserved) value of $H_0$ is equal
to
\begin{equation}\label{eq:24a}
H_0\approx -\bar{A}N_{k}^2(0)\ll E_0.
\end{equation}
For an arbitrary time, Eq. (\ref{eq:24}) can be rewritten as
$2E_0\sin\varphi_{k}(t)=-[2\xi_k+\bar{A}N_{k}(t)]-H_0/N_{k}(t)$.
Substituting this relation into equations (\ref{eq:21}), one can
find
\begin{equation} \label{eq:25}
\frac{dN_{k}}{dt}=\pm 2N_{k}\alpha;\;\;\;
\frac{d\varphi_{k}}{dt}=-\bar{A}N_{k}-\frac{H_0}{N_{k}},
\end{equation}
$\alpha=[E^2_0 -(\bar{A}N_{k}/2+\xi_k-H_0/2N_{k})^2]^{1/2}$. The
signs $\pm$ correspond to the regions with
$|\varphi_{k}|\leq\pi/2$ and $|\varphi_{k}|>\pi/2$, respectively.
For the region with $|\varphi_{k}|\leq\pi/2$, the solution can be
found straightforwardly
\begin{equation}\label{eq:27}
2t\approx \int\limits^{N_{k}(t)}_{N_{k}(0)}\frac{dx}{x[E^2_0
-(\bar{A}x/2+\xi_k)^2]^{1/2}},
\end{equation}
Here we omitted the small term with $|H_0|\ll E_0$. The upper
limit of the integral is equal to the value $N_{k}^m\approx
2(E_0\mp\xi_k)/|\bar{A}|$ at which the denominator vanishes. The
divergence at $N_k\rightarrow 0$ is a typical manifestation of the
parametric resonance which requires a nonzero initial field
amplitude. This can be achieved by taking  zero-point oscillations
into account \cite{K1}. Owing to these oscillations, we can put
$N_{k}(0)\sim 1$. The time necessary for the system to achieve the
maximal value $N_{k}^m$ is equal to
\begin{equation}\label{eq:27a}
t_m\approx
\frac{1}{2\alpha_{0k}}\ln\Big[\frac{4E_0}{\bar{A}}\cdot\Big(1-\frac{\xi_k^2}{E_0^2}\Big)\Big],
\end{equation}
We see that the argument of logarithm is much greater than unity
at $\xi_k<E_0$ and $E_0\gg \bar{A}$. This implies that $t_m\gg
1/2\alpha_{0k}$ where $1/2\alpha_{0k}$ is the characteristic time
of the parametric resonance. At $t=t_m$ we have
$|\varphi_{k}|=\pi/2$. As it follows from the second equation in
(\ref{eq:25}), at $t>t_m$ the phase proves to be in the region
with $|\varphi_{k}|>\pi/2$. As a result, in the first equation the
sign becomes negative and the phonon number reduces. This is the
start of an oscillating behaviour. The numerical simulation shows
that the joint solution of Eqs. (\ref{eq:25}) has the form of
anharmonic oscillations of both the phonon number and the phase
around their stationary values. At $\xi_k=0$ the exact solution
for $N_{k}(t)$ can be written in terms of dn Jacobian elliptic
function as
\begin{equation}\label{eq:IV.5}
N_{k}(t)=N_{k}^m
dn[2E_0(t-t_m);k],\;\;\;N_{k}^m=\frac{2E_0}{|\bar{A}|}.
\end{equation}
Where $dn(u;k)=dn(-u;k)$, $dn(0;k)=1$; $k$ is the elliptic modulus
\cite{A}. For the given initial conditions we have
$\sqrt{1-k^2}=|H_0 \bar{A}|/4E_0^2\ll 1$.

{\bf 3.} The character of evolution changes drastically at
$\gamma\neq0$. Now in all cases the system asymptotically
approaches the stationary state. At any set of initial parameters
the solution of Eq. (\ref{eq:21}) leads to the values
(\ref{eq:23}) for these stationary states. The damping of the
oscillations has the decrement, which is close to $\gamma$, when
$\alpha_0\gg\gamma$ and $t>t_m$. As $\gamma$ increases, the
arrival time to the stationary state decreases as $1/\gamma$. When
$\gamma$ is relatively close to $\alpha_{0k}$, the oscillations
disappear.  In this case the time necessary to reach the
stationary state is allied to the value (\ref{eq:27a}), if
$\alpha_{0k}$ is replaced by $\alpha_1=\alpha_{0k}-\gamma$.

\vspace{0.5cm}
\begin{figure}[!h] \centering\vspace*{-0.5cm}
  \begin{minipage}[t]{.24\textwidth} \centering
    \includegraphics[angle=-90, width=\textwidth]{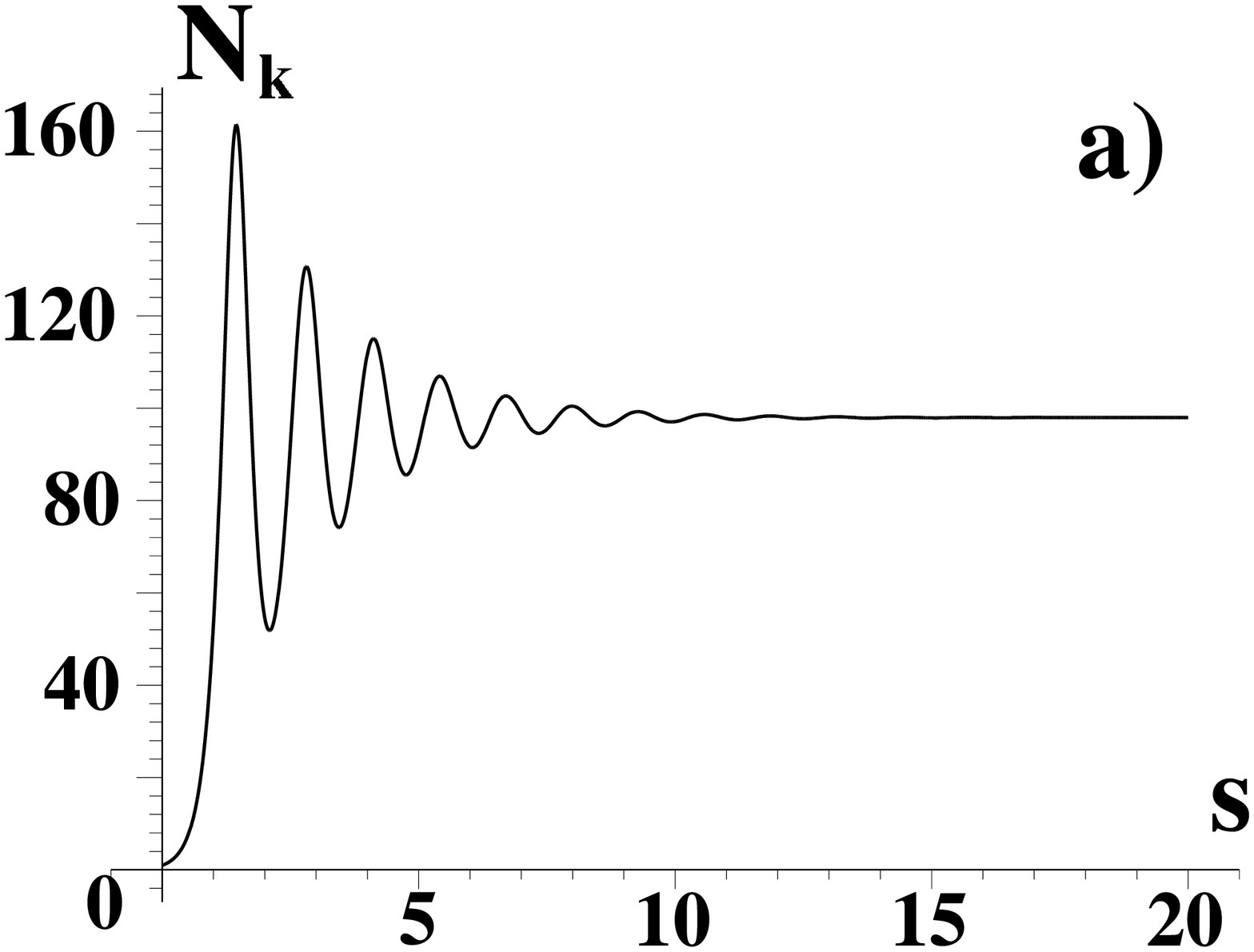}
\end{minipage}
  \begin{minipage}[b]{.23\textwidth} \centering
 \includegraphics[angle=-90, width=\textwidth]{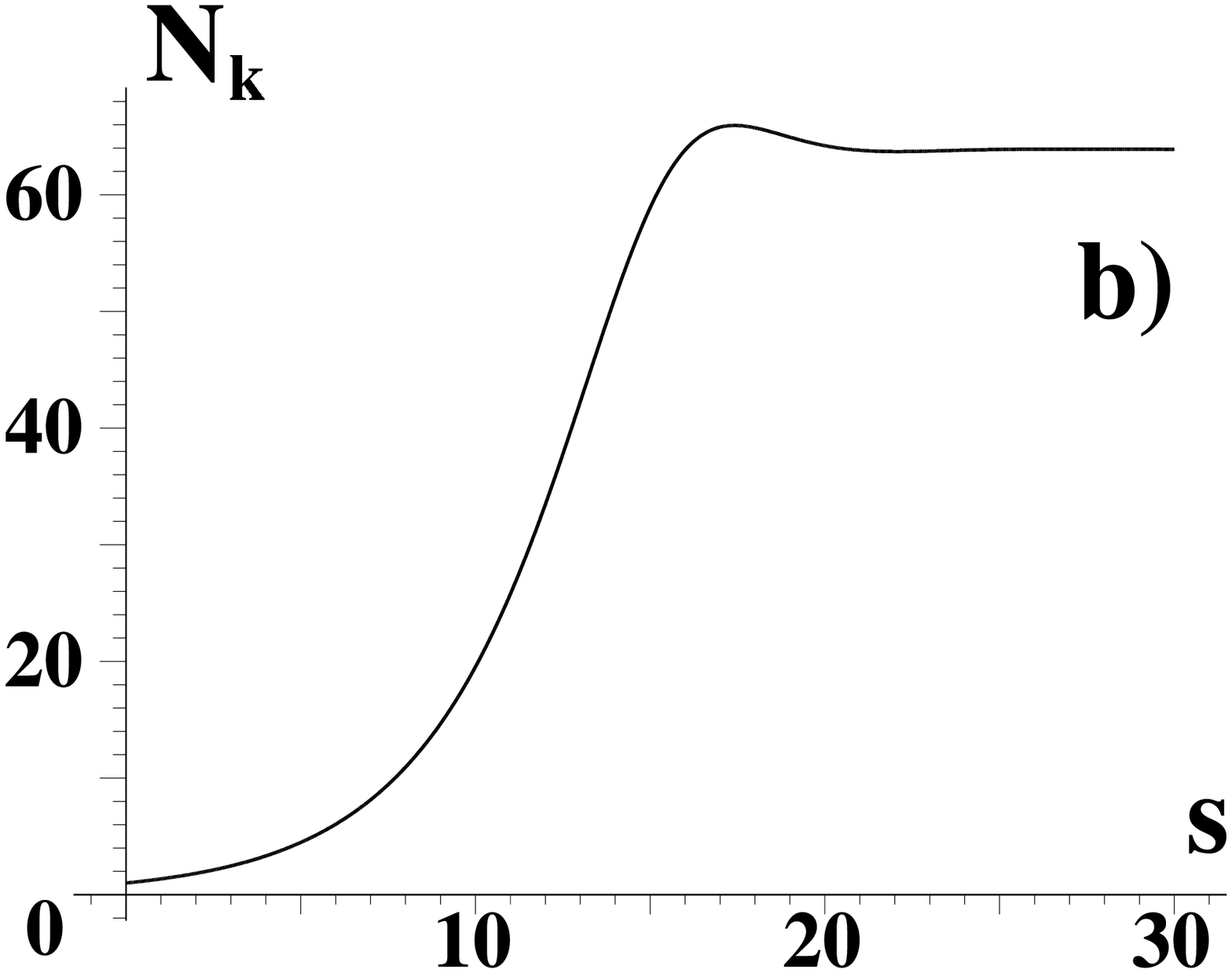}
\end{minipage}
\end{figure}

\begin{figure}[!h] \centering\vspace*{-0.5cm}
  \begin{minipage}[t]{.225\textwidth} \centering
    \includegraphics[angle=-90, width=\textwidth]{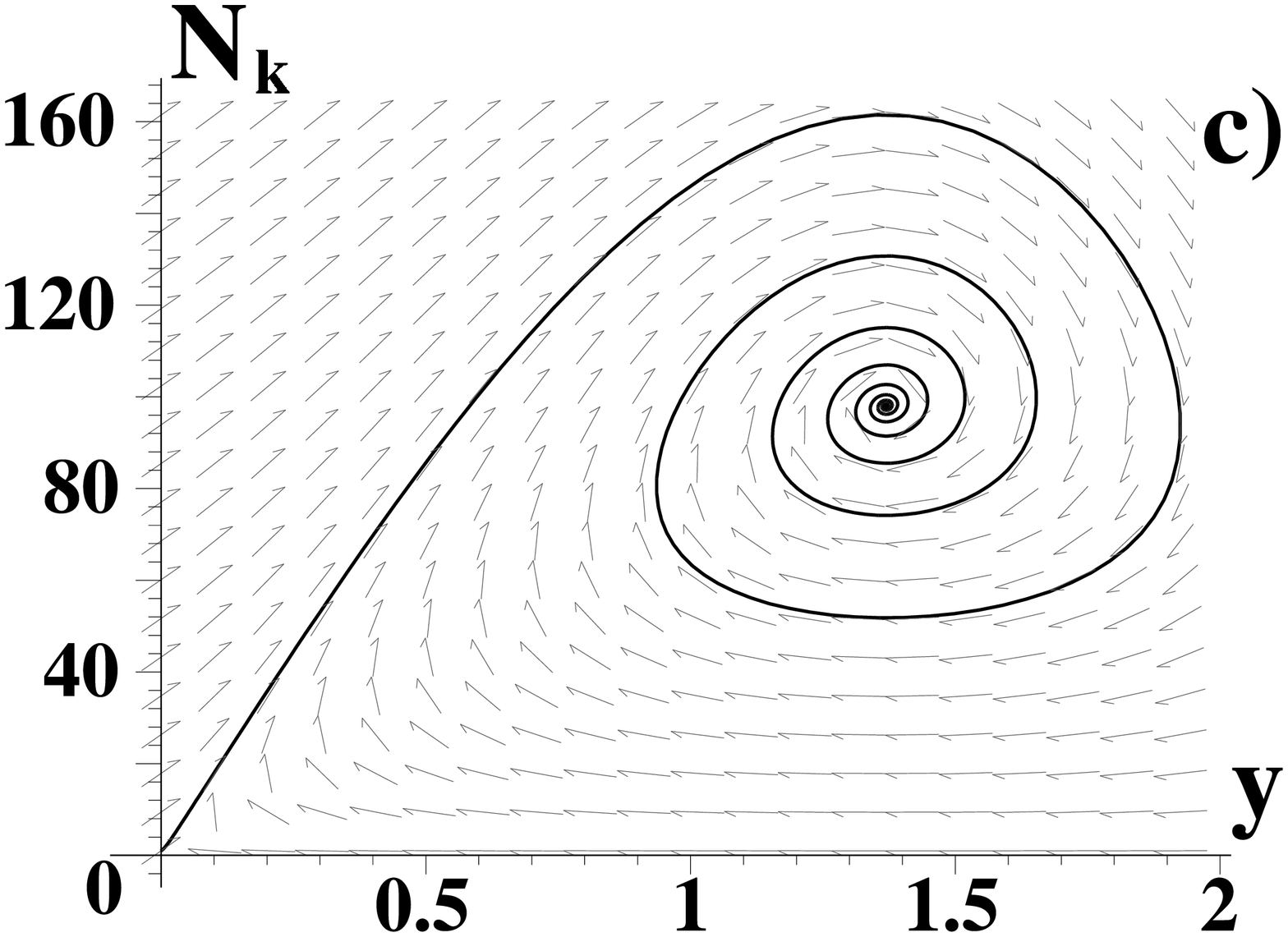}
\end{minipage}
  \begin{minipage}[b]{.225\textwidth} \centering
 \includegraphics[angle=-90, width=\textwidth]{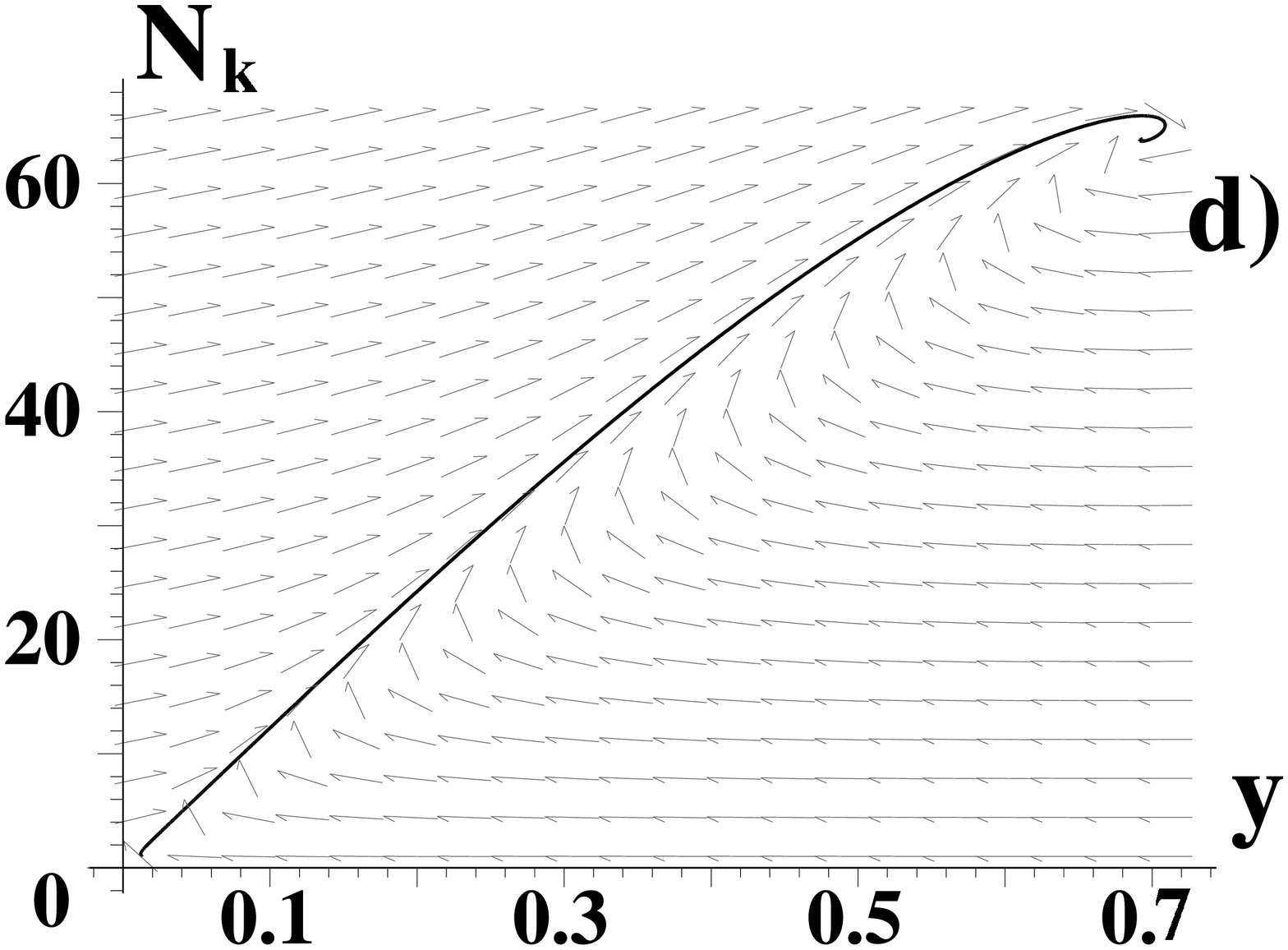}
\end{minipage}
\caption{Phonon numbers $N_k(s)$ versus $s=2\gamma t$ (a,b) and
versus $y(t)\equiv -\varphi(t)$ (c,d) for $(E_0/\gamma)=5,1.3$.}
\end{figure}

The direct numerical simulation of the system (\ref{eq:21})
demonstrates the picture described. As an example, in Fig.1 we
report the results for the phonon number $N_{k}$ as a function of
the dimensionless parameter $s=2\gamma t$ for $E_0/\gamma=5$ (a)
and $E_0/\gamma=1.3$ (b). Figs. 2 c), d) demostrate the phase
portraits in the plane $[N_{k},\varphi_{k}]$ for these cases. (The
parameters $(E_0/\bar{A})=10^2$ and $\xi=0$ are used for the
calculations.) In the both cases $N_{k}(t),\varphi_{k}(t)$ are
assymptotically approaching to the values (\ref{eq:22}).

Thus at any finite values of $\gamma$ the self-consistent
evolution of interacting phonons experienced the parametric
amplification results in the formation of the stationary state
with the macroscopic number (of the order of $E_0/\bar{A}$) of
phonons in a single state.

It should be noted that the character of the temporal evolution is
sign-independent of $\bar{A}$.

It is interesting to verify the stability of the solution
obtained. It seems that the damping of fluctuations of the phonon
number  $\delta N=N_{k}-N_{k}^s$  may have a different character
within and beyond the parametric interval. Really, at $\delta N<0$
the phonon number reverts to the stationary state owing to
parametric pumping. While at $\delta N>0$, the absorption alone,
determined by $\gamma$, returns the phonon system to this state.

Linearizing Eqs. (\ref{eq:21}) in $\delta N$ and $\delta
\varphi=\varphi_{k}-\varphi_{k}^s$ with the initial conditions
$\delta N(0)=N_{k}(0)-N_{k}^s$ and $\delta \varphi(0)=0$, we
obtain
\begin{equation} \label{eq:V.3}
\begin{split}
&\frac{d\delta N}{dt}=-[2E_0 N_{k}^s \sin\varphi_{k}^s] \delta
\varphi;\\ & \frac{d\delta \varphi}{dt}=-2\bar{A}\delta
N-[2E_0\cos\varphi_{k}^s]\delta \varphi,
\end{split}
\end{equation}
where $N_{k}^s$ and $\sin\varphi_{k}^s$ are determined by Eqs.
(\ref{eq:22}). Taking into account that
$\bar{A}\sin\varphi_{k}^s<0$ for the both signs of $\bar{A}$, we
find the solution of Eqs. (\ref{eq:V.3a}) in the form

\begin{equation} \label{eq:V.3a}
\begin{split}
&\delta N(t)=\delta N(0)e^{-\gamma t}\Big(\cos\omega
t-\frac{\gamma}{\omega}\sin\omega t\Big);\\ &\delta
\varphi(t)=\frac{-2\delta N(0)e^{-\gamma t}|\bar{A}|}
{\omega}\cdot\sin\omega t\\ &
\omega^2=4\Big(E_0^2-\gamma^2\Big)^{1/2}\cdot\Big[\Big(E_0^2-\gamma^2\Big)^{1/2}-\xi_k\Big]-\gamma^2.
\end{split}
\end{equation}
This result demonstrates the stability of the stationary state
against fluctuations of the phonon density. The character of
relaxation to the stationary state turns out sign-independent of
$\delta N(0)$.

The quantitave character of the solution with the damping
fluctuations remains for the case $\delta\varphi(0)\neq 0$.

\section{V. Solution for the case of several levels}
{\bf 1.} Let us consider the case when several discrete levels lie
within the parametric interval. First of all, we analyze in detail
the evolution to the stationary state for two levels at
$\gamma\neq 0$. In the case of two levels, taking into account
double degenaracy of each level in Eq. (\ref{eq:C1}) and
introducing notations
$N_{k_i}=N_{-k_i}=N_i;\;\varphi_{k_i}=\varphi_{-k_i}=\varphi_i;\;i=1,2$,
we obtain
\begin{equation} \label{eq:C5}
\begin{split}
&\!\!\!\!\!\!\frac{dN_{1}}{dt}\!=\!\!-2\gamma N_{1}\!+\!2E_0
N_{1}\cos\varphi_{1}\!\!-4AN_{1} N_{2}\sin(\varphi_{1} \!\!
-\varphi_{2});\\ &\!\!\!\!\!\!\frac{d\varphi_{1}}{dt}=-2\xi_1-2E_0
\sin\varphi_{1}-6AN_{1}-8AN_{2}-\\ &
\;\;\;\;\;\;\;\;\;\;\;\;\;\;\;\;\;\;\;\;\;\;\;\;\;\;\;\;\;\;\;\;\;\;\;\;\;\;\;\;\;\;\;-4AN_{2}\cos(\varphi_{1}-\varphi_{2});\\
&\!\!\!\!\!\!\frac{dN_{2}}{dt}\!=\!\!-2\gamma N_{2}\!+\!2E_0
N_{2}\cos\varphi_{2}\!+\!4AN_{1}
N_{2}\sin(\varphi_{1}\!\!-\varphi_{2});\\
&\!\!\!\!\!\!\frac{d\varphi_{2}}{dt}=-2\xi_2-2E_0
\sin\varphi_{2}-6AN_{2}-8AN_{1}-\\ &
\;\;\;\;\;\;\;\;\;\;\;\;\;\;\;\;\;\;\;\;\;\;\;\;\;\;\;\;\;\;\;\;\;\;\;\;\;\;\;\;\;\;\;-4AN_{1}\cos(\varphi_{1}-\varphi_{2});
\end{split}
\end{equation}
Here the position of the renormalized level is determined by the
expression (see (\ref{eq:20a}))
\begin{equation} \label{eq:C2}
\bar{\xi}_1=\xi_1+3AN_1+4AN_2;\;\;\xi_1=\omega_1-\frac{\omega_0}{2};
\end{equation}
and, correspondently, for $\bar{\xi}_2$ with the replacement
$1\rightleftarrows 2$. The initial conditions are similar to those
in the case of a single level, namely, $N_{1,2}(0)\sim
1\;\;\;\;\sin\varphi_{i}(0)\approx-\xi_i/E_0$.

The solution of Eqs. (\ref{eq:C5}) demonstrates an interesting
asymptotic picture. For any set of the parameters in Eqs.
(\ref{eq:C5}), the number of phonons corresponding to one of two
levels vanishes. Owing to renormalization determined by Eq.
(\ref{eq:C2}), another level proves to be at the edge of the
parametric interval with the number of phonons equal to $N^s_i$
from Eq.(\ref{eq:22}). \vspace{0.5cm}
\begin{figure}[!ht] \centering\vspace*{-0.5cm}
  \begin{minipage}[t]{.23\textwidth} \centering
    \includegraphics[angle=-90, width=\textwidth]{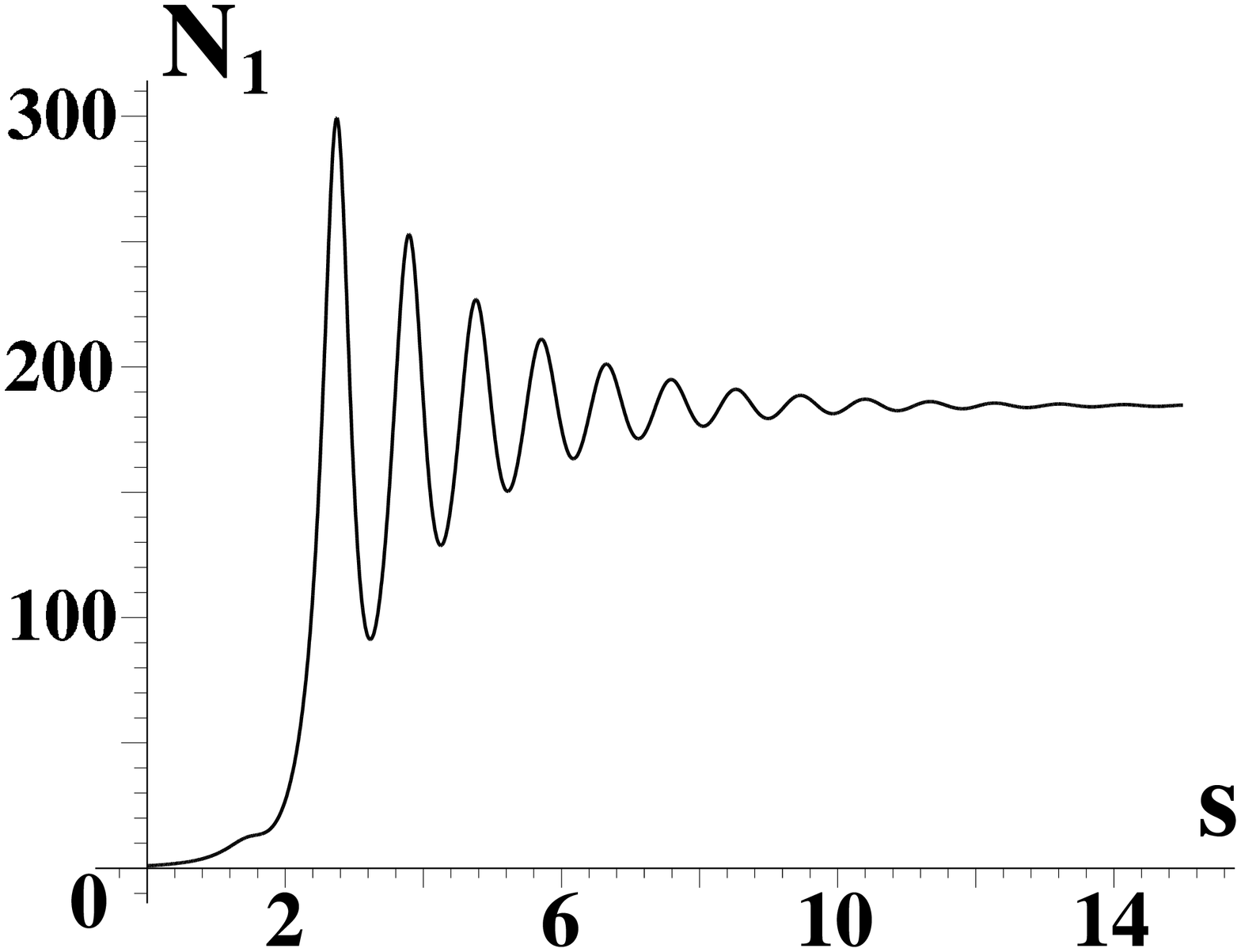}
\end{minipage}
  \begin{minipage}[b]{.234\textwidth} \centering
 \includegraphics[angle=-90, width=\textwidth]{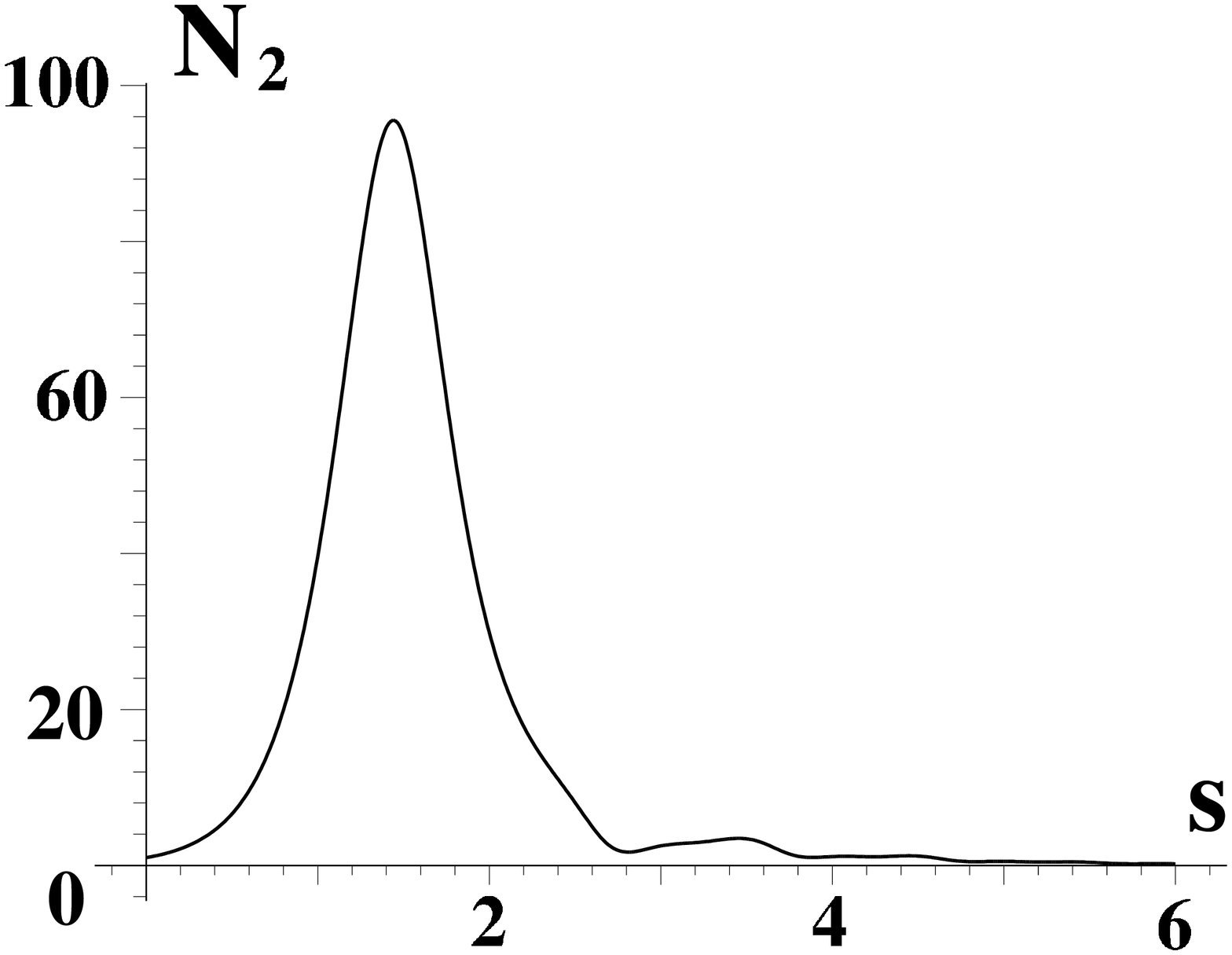}
\end{minipage}
\caption{Phonon numbers $N_{1,2}$  versus $s=2\gamma t$ for
$\xi_1/E_0=-0.87$, $\xi_2/E_0=0.3$ and $E_0/\gamma=5$.}
\end{figure}

\begin{figure}[!ht] \centering\vspace*{-0.5cm}
  \begin{minipage}[t]{.23\textwidth} \centering
    \includegraphics[angle=-90, width=\textwidth]{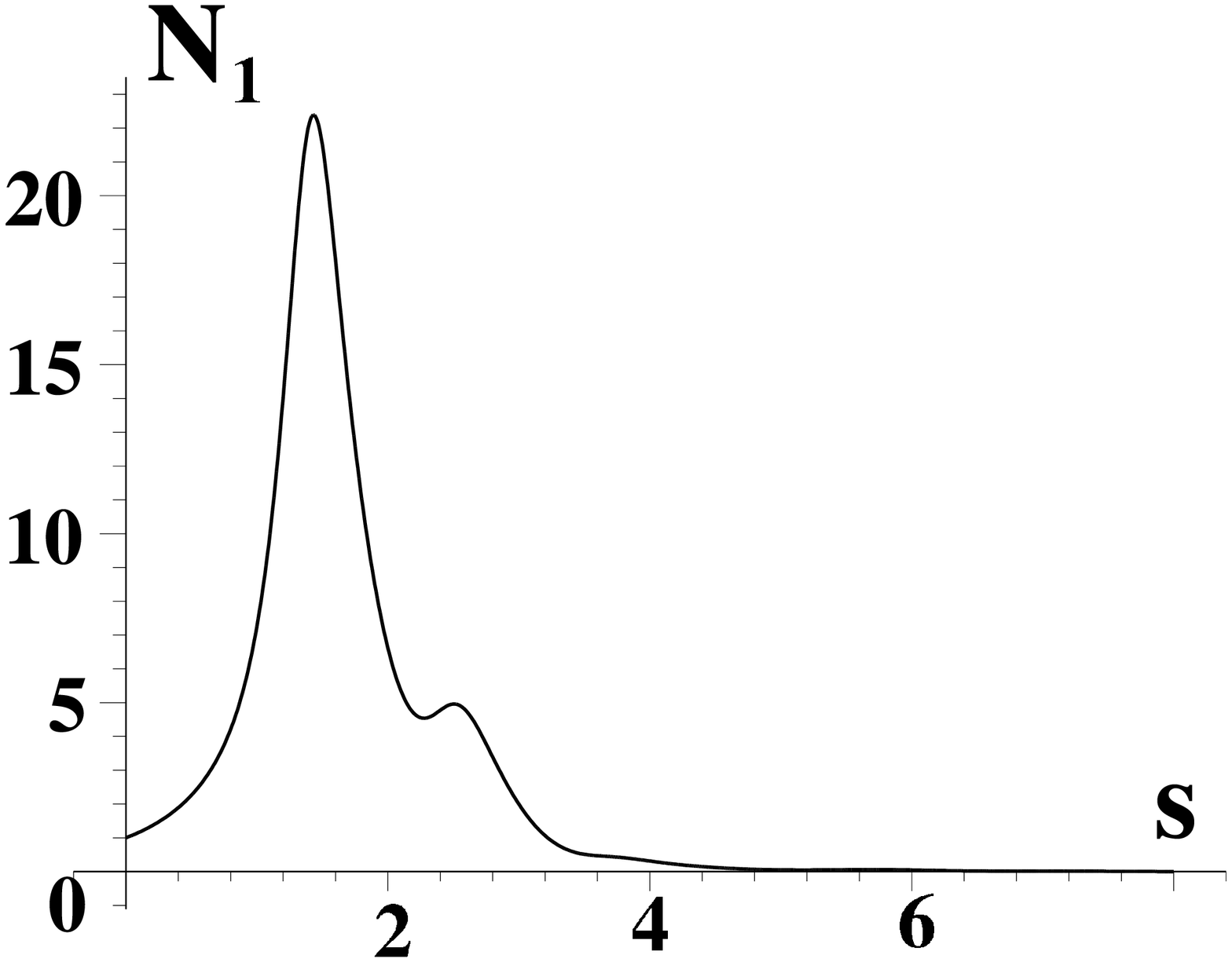}
\end{minipage}
  \begin{minipage}[b]{.234\textwidth} \centering
 \includegraphics[angle=-90, width=\textwidth]{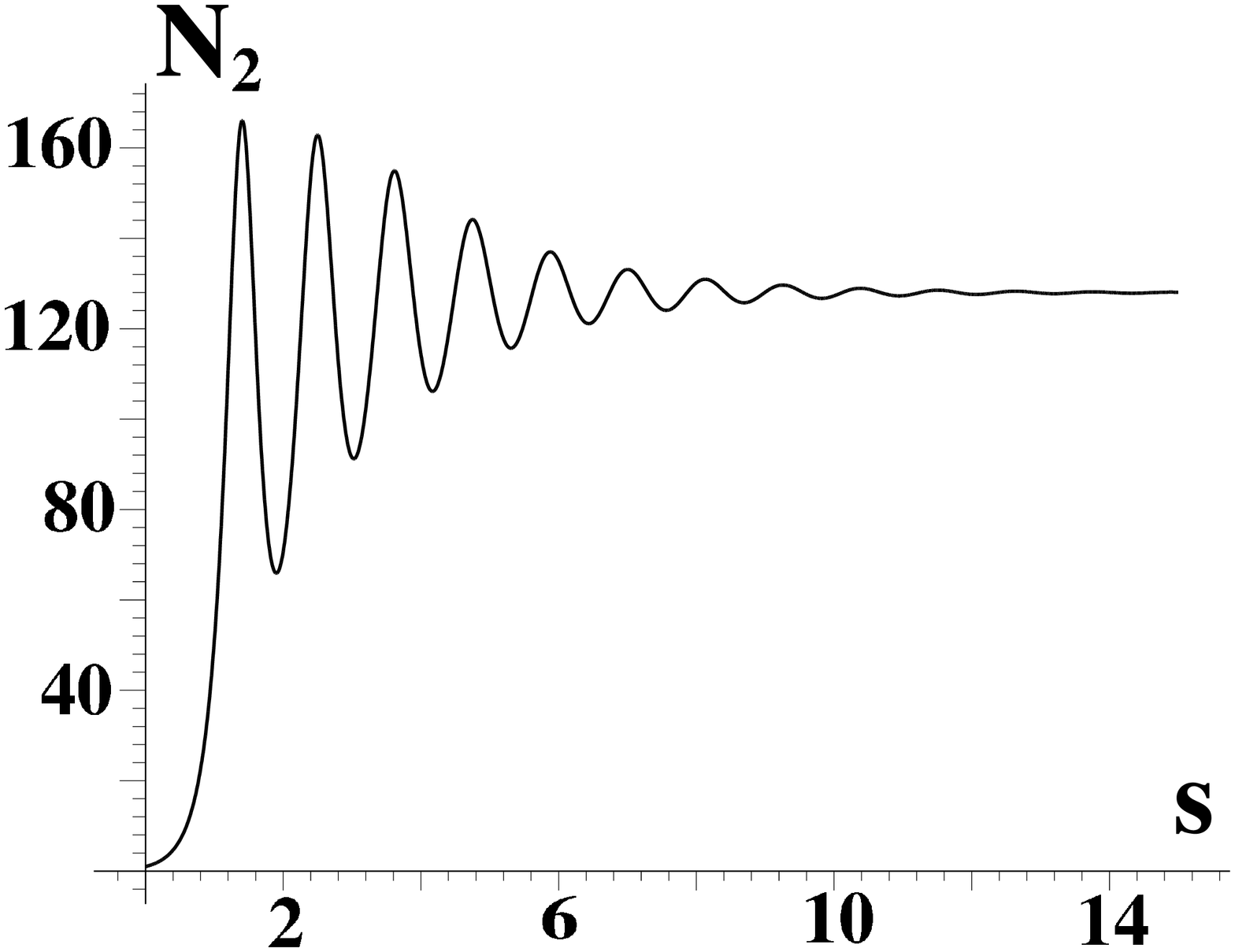}
\end{minipage}
\caption{Phonon numbers $N_{1,2}$ versus $s=2\gamma t$ for
$(E_0/\gamma)=5$ and $(\xi_1/E_0)=-0.87$, $(\xi_2/E_0)=-0.3$.}
\end{figure}

The disappearance of phonons at one of the levels is due to the
following. This level, as a result of renormalization, proves to
be either beyond the parametric region in which pumping is absent,
or within the interval beside its edge where the damping exceeds
the resonance pumping. In the both cases the growth of the phonon
occupation numbers is suppressed by the damping.

In most cases we have the characteristic picture represented in
Fig.2 when the level 1 closer to the left edge of the interval at
the initial time moment ("the left level") reaches the stationary
state ($E_0, \bar{A}>0$).

It is interesting that, owing to renormalization, the level 2
closer to the right edge of the parametric interval ("the right
level") asymptotically proves to be far from the edge,
$\bar{\xi}_2/E_0=2.75$.

Figure 3 demonstrates another characteristic situation. Now the
right level $\xi_2$ comes to the stationary state. The
renormalization, due to $N_{2}$, shifts the level $\xi_1$ to the
right edge of the parametric interval, $\bar{\xi}_1/E_0=0.84$,
where the subsequent evolution of the moderate values of $N_{1}$
is suppressed by $\gamma$.

\vspace{0.5cm}
\begin{figure}[!ht] \centering\vspace*{-0.5cm}
\begin{minipage}[t]{.43\textwidth} \centering
\includegraphics[angle=360, width=\textwidth]{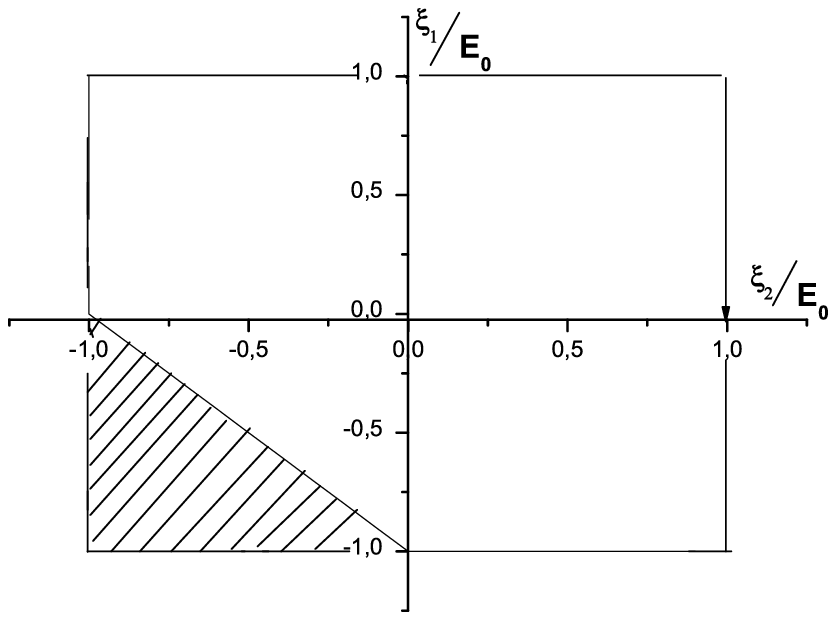}
\end{minipage}
\caption{In the shaded region the right level comes to the
stationary state. In the rest part of the plane $[\xi_1;\;\xi_2]$
the left level arrives at this state($E_0/\gamma=5$)}
\end{figure}
Figure 4 displays two regions in the plane [$\xi_1$, $\xi_2$] for
the fixed value $(E_0/\gamma)=5$. For the positions $\xi_1$,
$\xi_2$ from the shaded region, the right level $\xi_2$ arrives at
the stationary state. For the other positions, the left level
$\xi_1$ reaches this state.

The essential remarks should be added. In principle, owing to the
renormalization,  an outer level with $|\xi_1|/E_0>1$ ($\xi_1< 0$)
can be indrawn into the parametric interval. It is impossible for
any ratio $E_0/\gamma$, if $\xi_2< 0$. This is already clear from
Fig.4 (with decreasing $E_0/\gamma$ the shaded region only
enlarges).

\begin{figure}[!ht] \centering\vspace*{-0.5cm}
\begin{minipage}[t]{.45\textwidth} \centering
\includegraphics[angle=360, width=\textwidth]{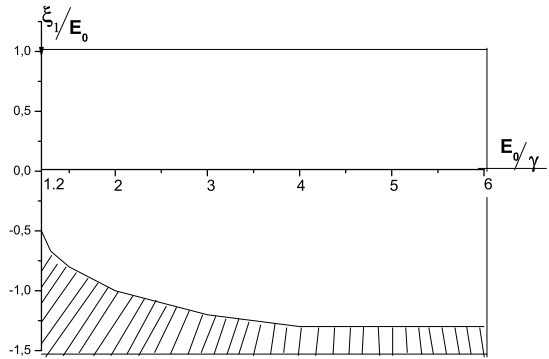}
\end{minipage}
\caption{The shaded region covers the level positions at the fixed
ratio $\xi_2/E_0=0.4$, including the levels with $\xi_1/E_0< -1$,
which are damped in the course of the evolution. The stationary
state is approached by the state with $\xi_2$.}
\end{figure}

However, it is possible, when $\xi_2> 0$. Moreover, the indrawn
level can assymptotically approach  the edge of the parametrical
interval forming the stationary state. For any $\xi_2> 0$ there is
the limited interval $\delta_c(\xi_2)=(|\xi_{1c}|-E_0)/E_0$, when
this event is not realized. At $\delta<\delta_c$ the indrawn level
ousts the level $\xi_2$ and reaches the stationary position as
long as the ratio $E_0/\gamma$ exceeds some critical value. The
interval $\delta_c$ increases with decreasing $\xi_2$. The direct
numerical calculation gives $(\delta_c)_{max}\approx 0.6$.

As an illustration, in Fig.5 we plot the plane
$[\xi_1/E_0;\;E_0/\gamma]$ and distinguish two regions for the
fixed value $\xi_2/E_0=0.4$. In the shaded region the positions of
left levels, including  the ones with $\xi_1/E_0< -1$, which are
damped in the course of the evolution, are collected. In this case
the level with $\xi_2$ comes to the stationary state. The rest
part of the plane is occupied by the level positions (including
again indrawn levels at $\xi_1/E_0< -1$) for which the left level
ends the the evolution in the stationary state with the
macroscopic number of phonons (here we imply that $\xi_1<\xi_2$).

Actually, all features discussed here are represented in Fig.5.

\vspace{0.5cm}
\begin{figure}[!ht] \centering\vspace*{-0.5cm}
  \begin{minipage}[t]{.23\textwidth} \centering
    \includegraphics[angle=-90, width=\textwidth]{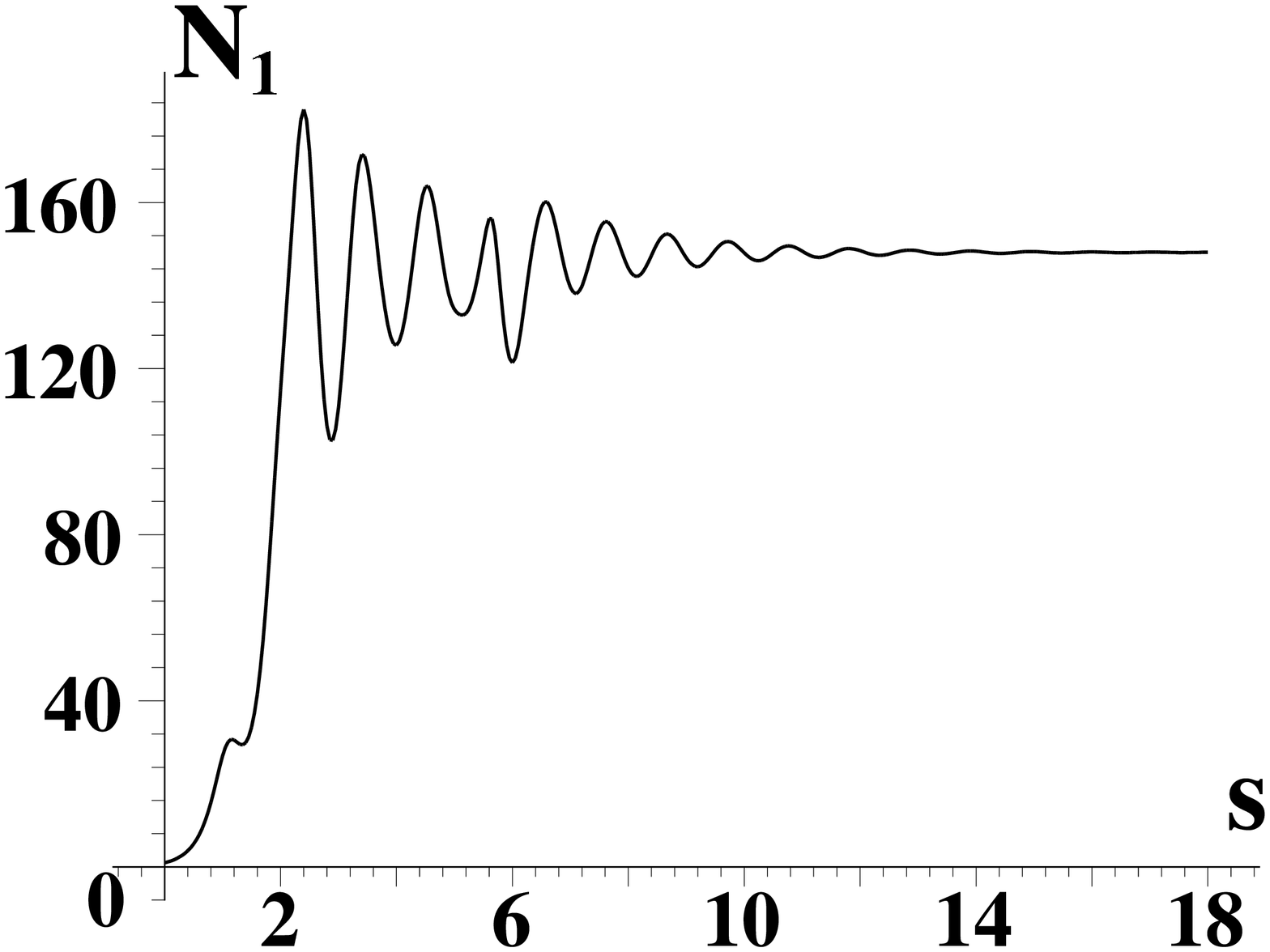}
\end{minipage}
  \begin{minipage}[b]{.23\textwidth} \centering
 \includegraphics[angle=-90, width=\textwidth]{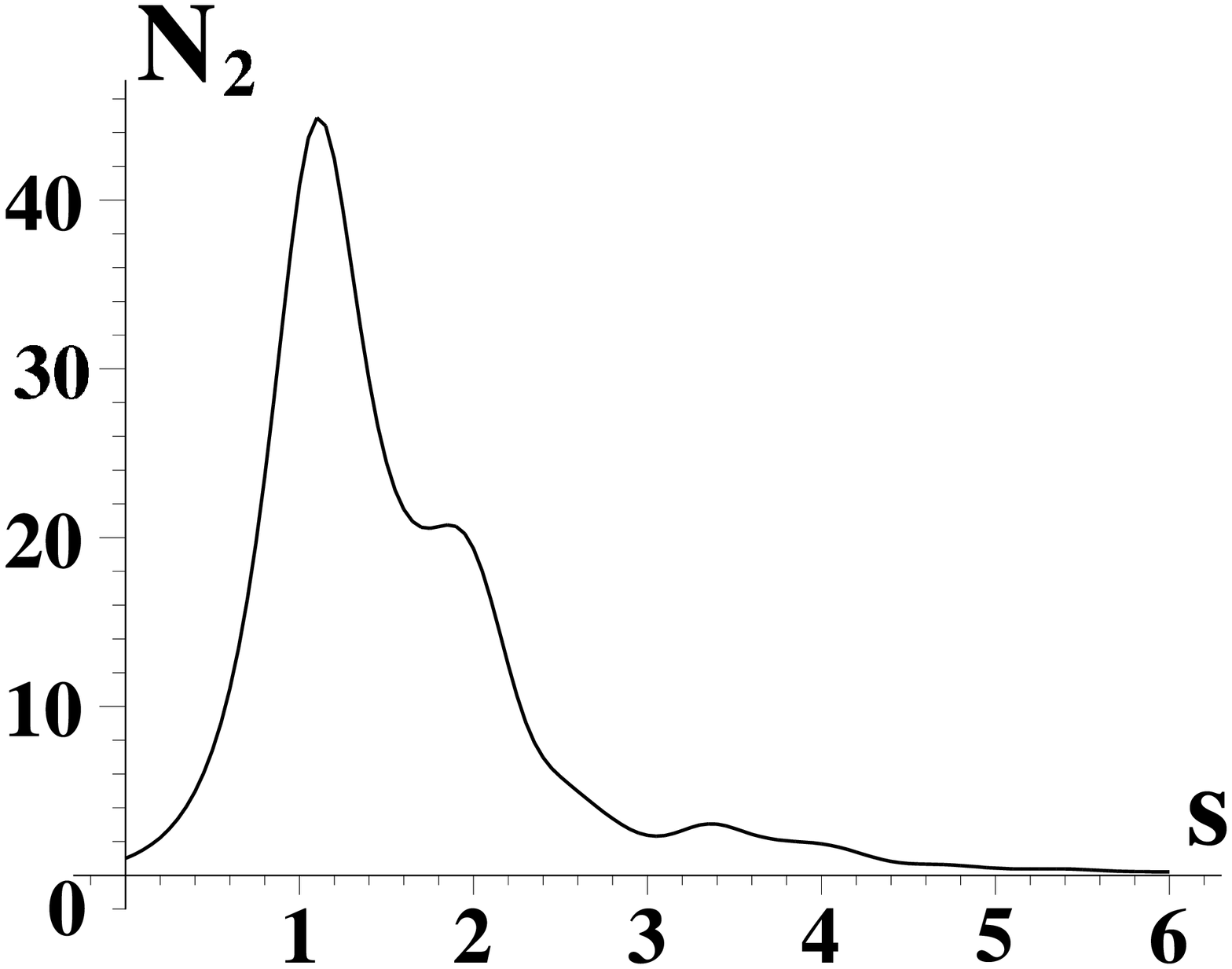}
\end{minipage}
\begin{minipage}[t]{.23\textwidth} \centering
    \includegraphics[angle=-90, width=\textwidth]{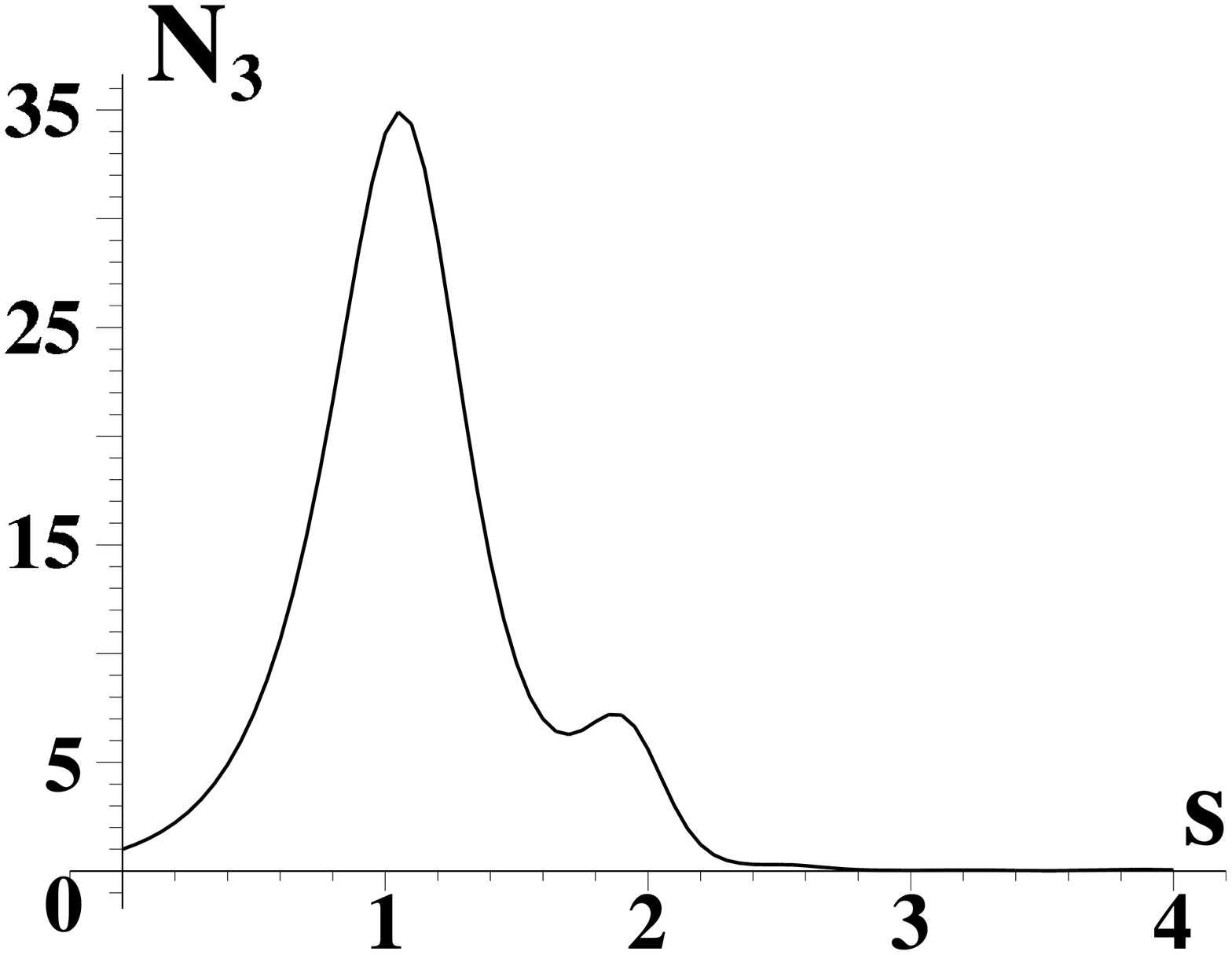}
\end{minipage}
\caption{Phonon numbers $N_{1,2,3}$  versus $s=2\gamma t$ for
$(\xi_1/E_0)=-0.5$, $(\xi_2/E_0)=-0.1$, $(\xi_3/E_0)=0.1$ and
$(E_0/\gamma)=5$.}
\end{figure}

{\bf 3.} All qualitative features revealed for two levels are
reproduced for the case of many levels within the parametric
interval. First of all, in the course of the evolution only one
level proves to be in the stationary state.The population of other
levels is going to zero in all cases. The picture of the evolution
as itself can demonstrate different patterns. In this aspect
already the case of three levels shown as an illustration in Figs.
6-8  is quite instructive. Really, in Fig.6 the leftmost level 1
is eventually found at the stationary state, while two other
levels are pushed out of the parametric interval,
$\bar{\xi}_2/E_0=1.83$, $\bar{\xi}_3/E_0=2.03$. \vspace{0.5cm}
\begin{figure}[!ht] \centering\vspace*{-0.7cm}
  \begin{minipage}[b]{.23\textwidth} \centering
 \includegraphics[angle=-90, width=\textwidth]{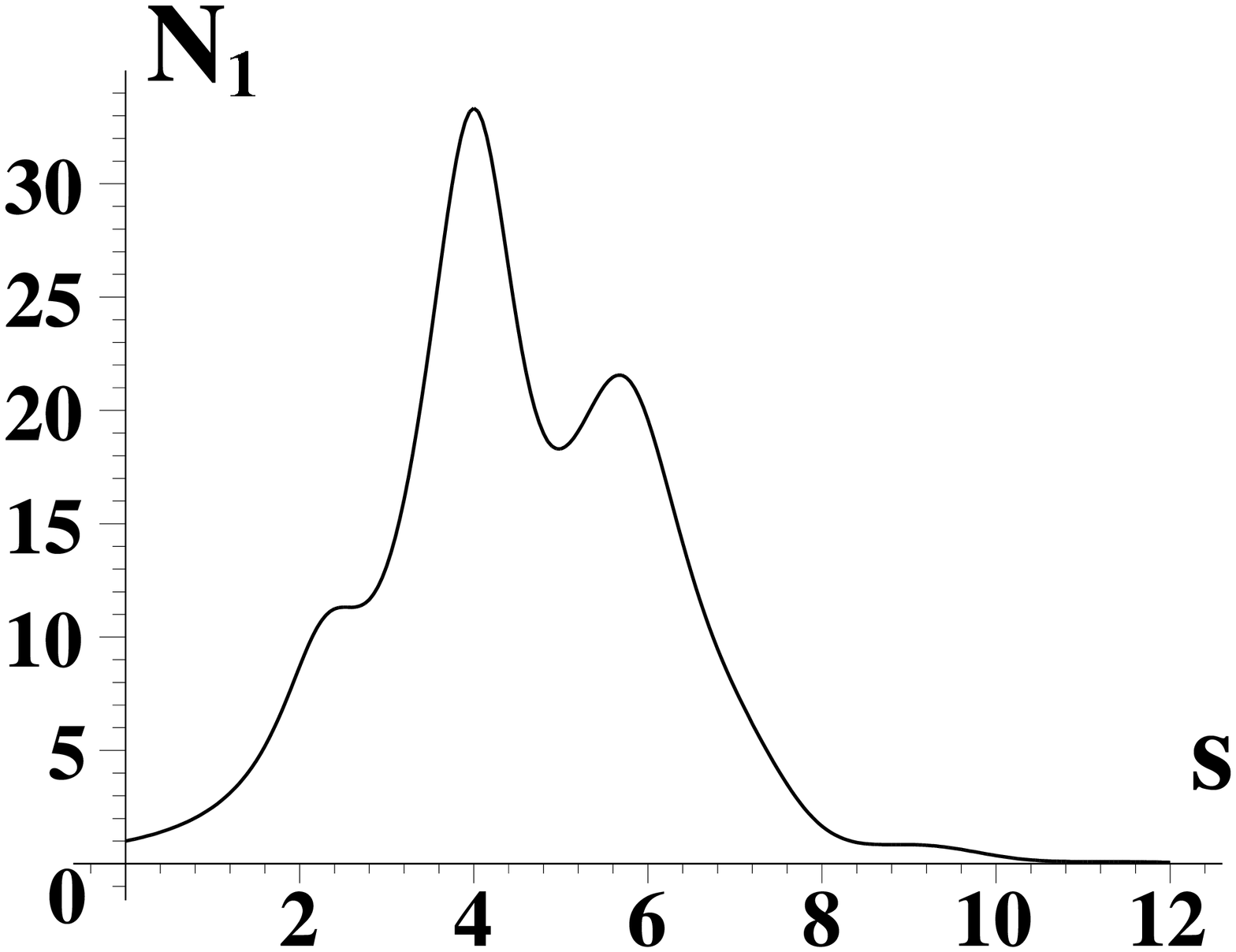}
\end{minipage}
  \begin{minipage}[t]{.23\textwidth} \centering
    \includegraphics[angle=-90, width=\textwidth]{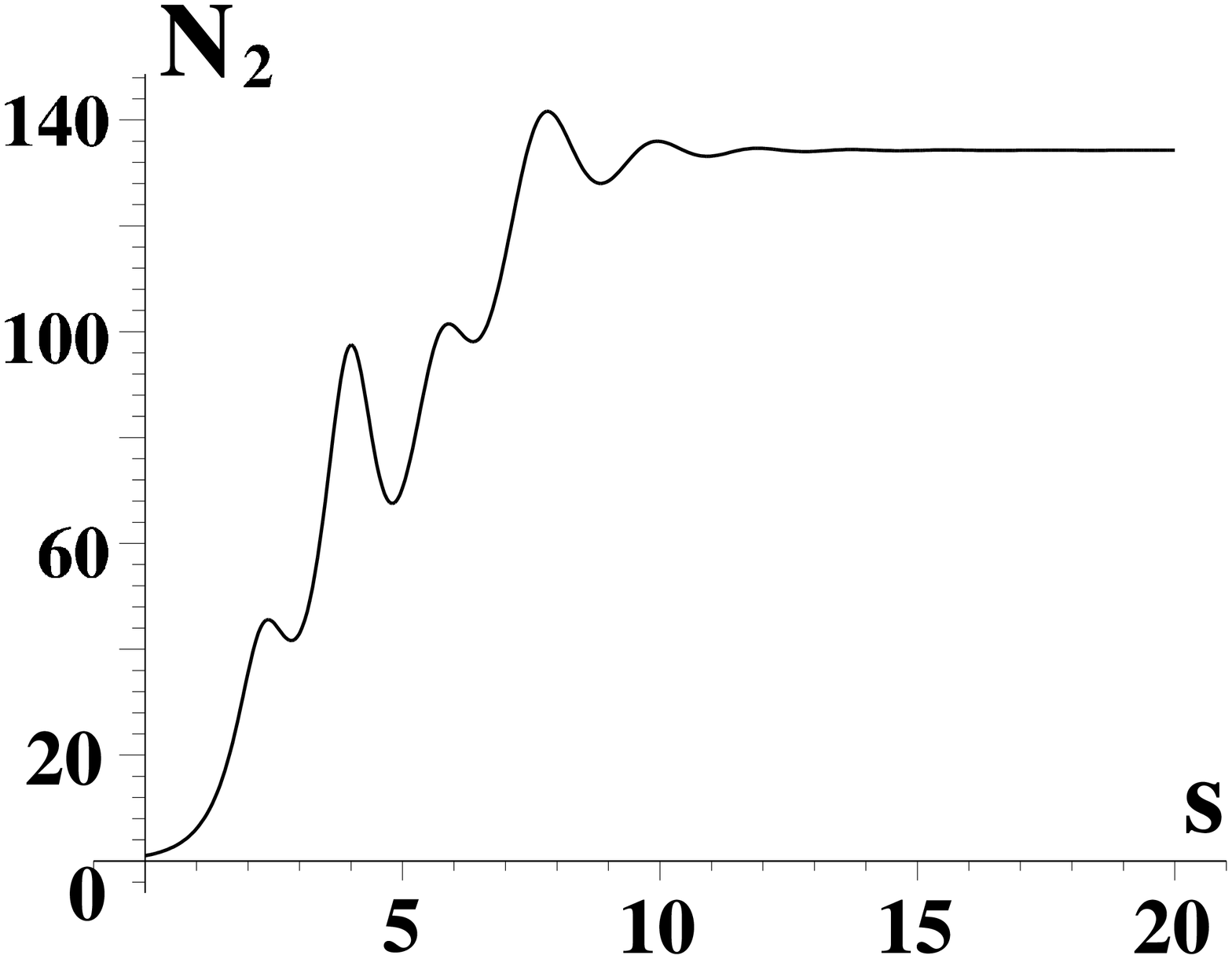}
\end{minipage}
  \begin{minipage}[b]{.23\textwidth} \centering
 \includegraphics[angle=-90, width=\textwidth]{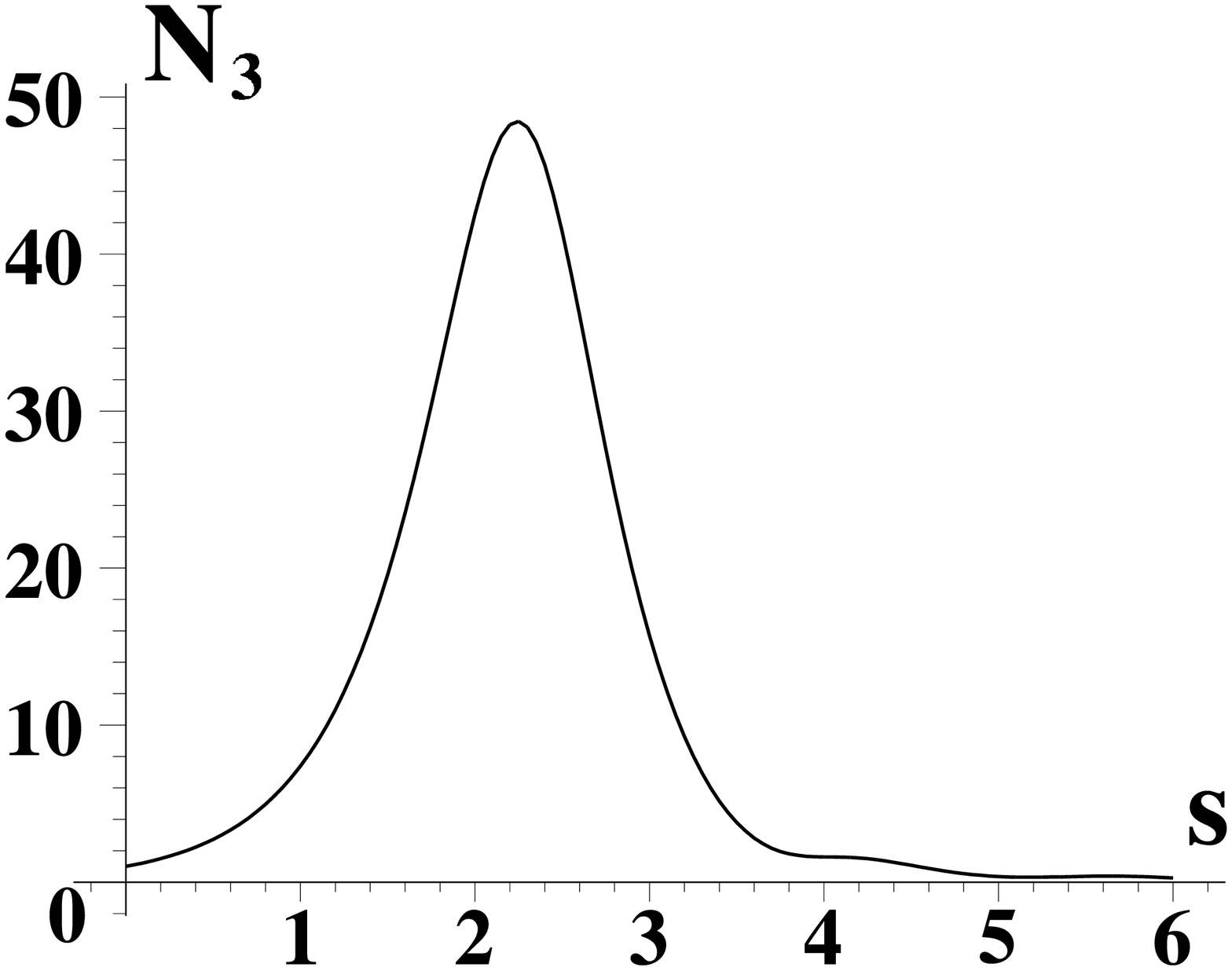}
\end{minipage}
\caption{{Phonon numbers $N_{1,2,3}$  versus $s=2\gamma t$ for
$(E_0/\gamma)=3$, $(\xi_1/E_0)=-0.8$, $(\xi_2/E_0)=-0.4$,
$(\xi_3/E_0)=0.1$.}}
\end{figure}

In Fig.7 the intermediate level 2 comes to the stationary state.
Now only the rightmost level is ousted from the parametric
interval ($\bar{\xi}_3/E_0=1.89$). The leftmost level 1 ends its
evolution with zero population at $\bar{\xi}_1/E_0=0.9$

\vspace{0.5cm}
\begin{figure}[!ht] \centering\vspace*{-0.5cm}
  \begin{minipage}[t]{.23\textwidth} \centering
    \includegraphics[angle=-90, width=\textwidth]{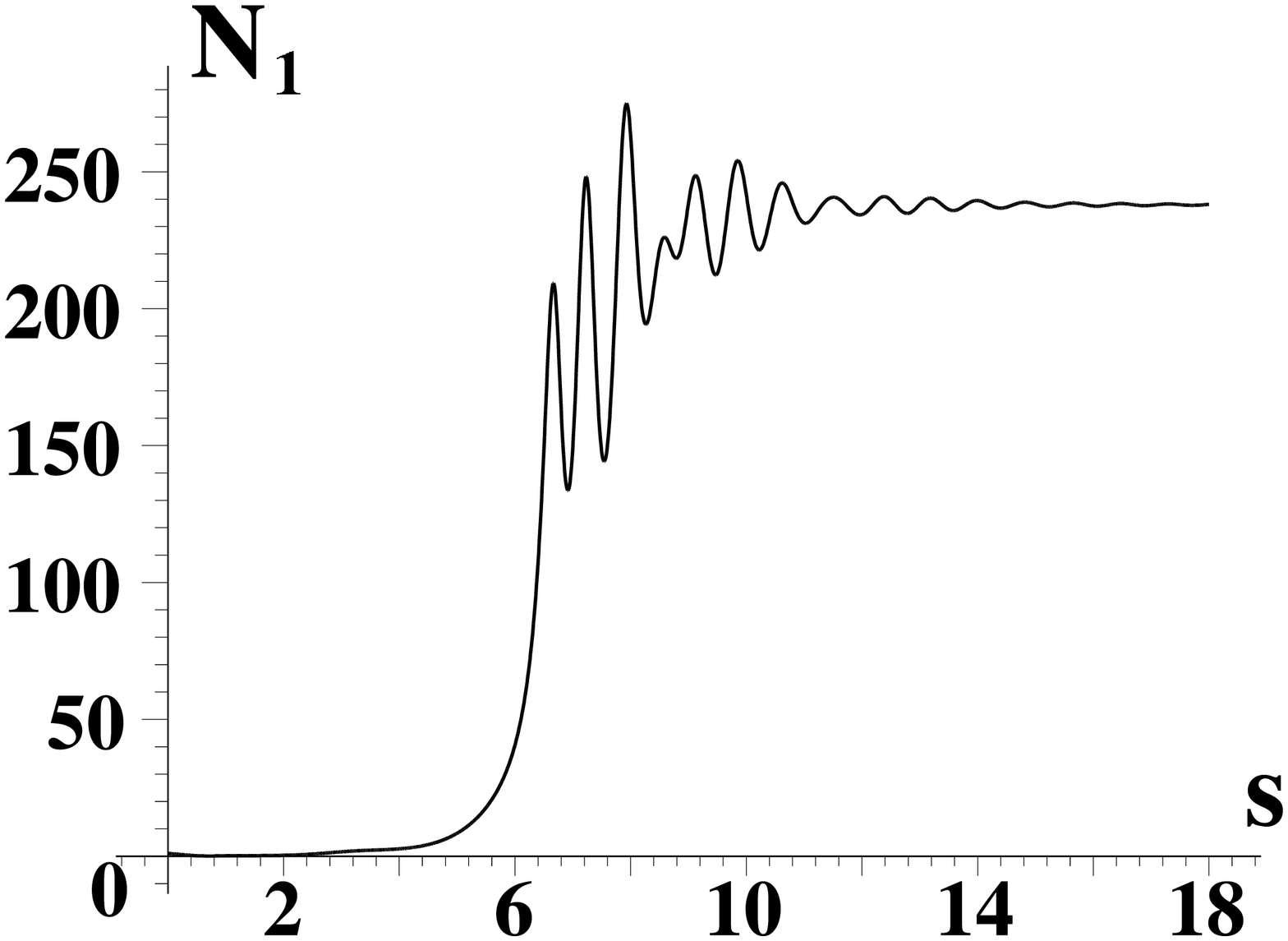}
\end{minipage}
  \begin{minipage}[b]{.23\textwidth} \centering
 \includegraphics[angle=-90, width=\textwidth]{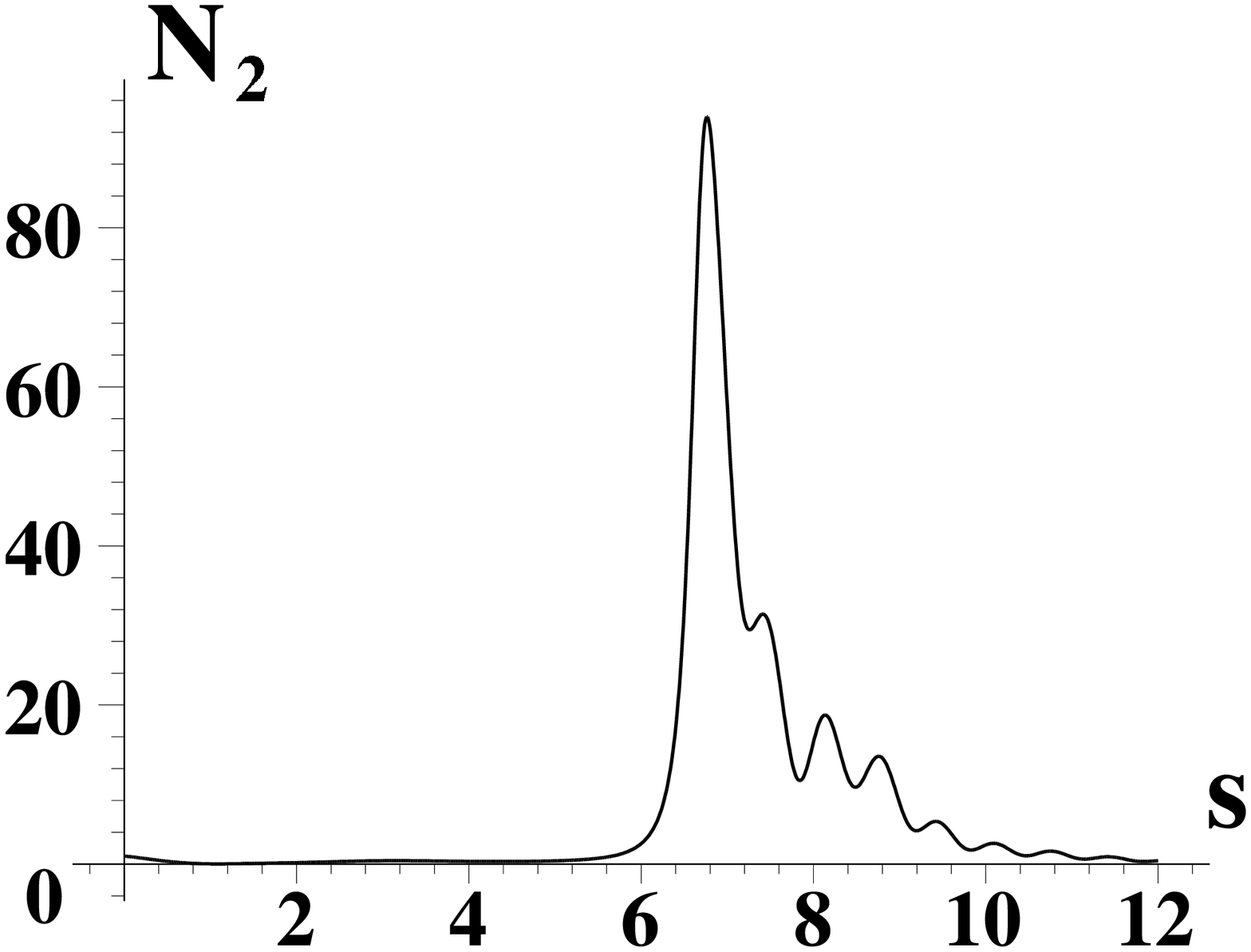}
\end{minipage}
\begin{minipage}[t]{.23\textwidth} \centering
    \includegraphics[angle=-90, width=\textwidth]{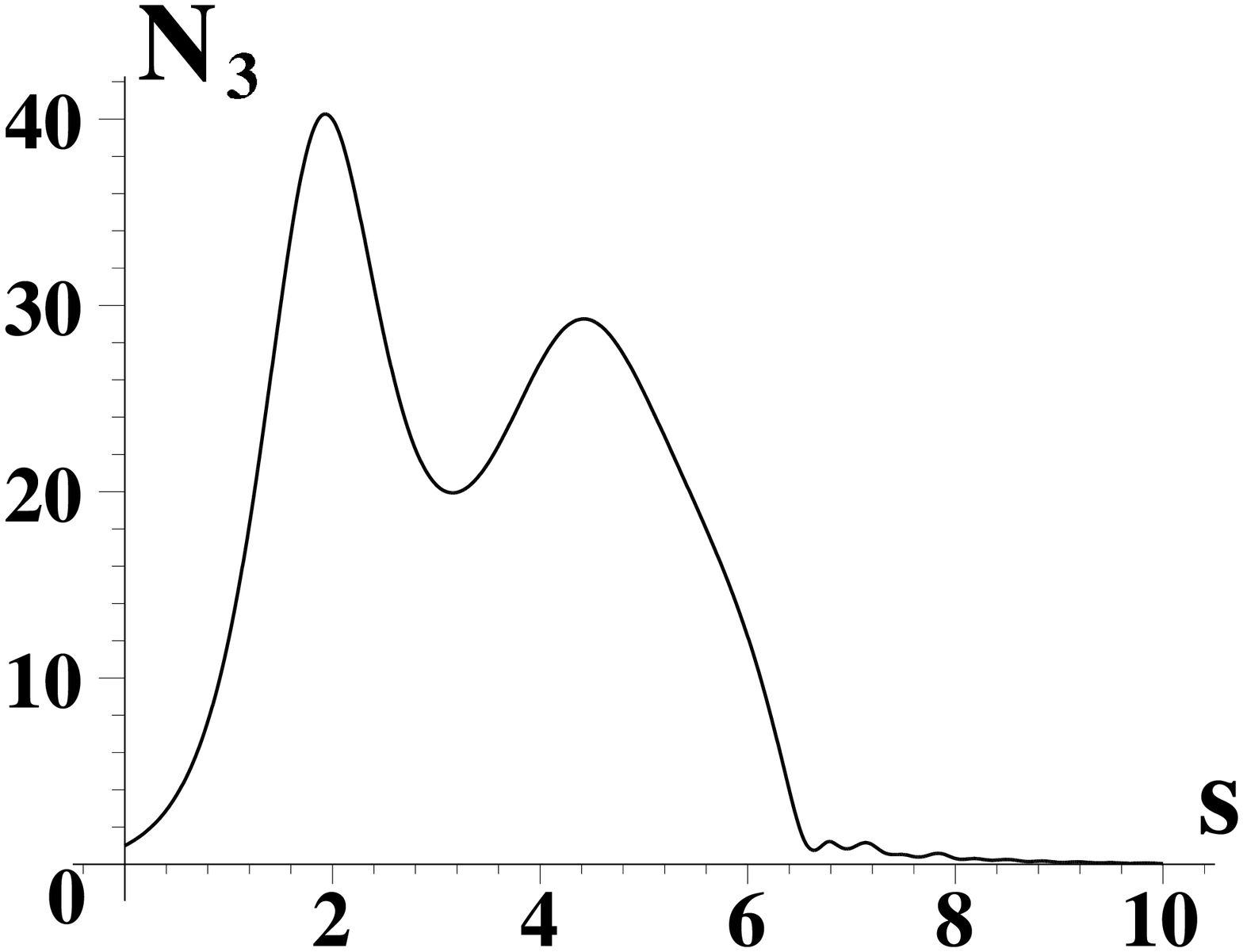}
\end{minipage}
\caption{{Phonon numbers $N_{1,2,3}$  versus $s=2\gamma t$ for
$(E_0/\gamma)=5$, $(\xi_1/E_0)=-1.4$, $(\xi_2/E_0)=-1.2$,
$(\xi_3/E_0)=0.7$.}}
\end{figure}

At last, Figure 8 displays the interesting situation when two
levels lie beyond the parametric interval, although they are
sufficiently close to the edge. In this case the both levels are
indrawn into the interval. The leftmost level arrives at the
stationary state.

It should be emphasized that in all cases the stationary state
arising in the course of the evolution has the number of phonons
and the phase equal to the values determined by the expressions
(\ref{eq:22}).

To conclude, we note that the character of the evolution described
above holds and for $A<0$. In this case the levels move to the
left edge of the parametric interval in the course of the
evolution.

\section {VI. Periodical stucture and some estimations}

{\bf 1.} The self-consistent temporal evolution of interacting
phonons near the parametric resonance results in the formation of
the stationary state with the macroscopic phonon number of about
$E_0/\bar{A}$ in a single quantum state with the common phase for
all excitations. The stationary state has one of two energies
corresponding to the right or left edge of the parametric
interval. In fact, the stationary state is formed by the phonon
pairs which are a coherent superposition of two states with the
wave vectors $k_s$ and $-k_s$ where $k_s=\omega_s/c_0$ and
$\omega_s=\omega_0/2\pm E_0$.

This state generates the stationary space modulation of the atomic
condensate. This is easy to reveal determining the atomic density
as $\delta n=<\hat{\chi}^{'+}({\bf \rho},z)\hat{\chi}'({\bf
\rho},z)>$, where $\hat{\chi}'$ has the form (\ref{eq:11}).
Rewriting $\hat{\chi}'$ in terms of phonon operators and
integrating over $d^2 \rho$, we arrive at the expression for the
modulation of the 1D atomic density
\begin{equation} \label{eq:29}
\frac{\delta
n(z)}{n^{(1)}}\!\!=\!\!\frac{2N^s_{k}}{N}(u_{k_s}^2+v_{k_s}^2)\cos
2k_s
z\!\!\approx\!\!\frac{2\mu}{\varepsilon_s}\cdot\frac{N_{k}^s}{N}\cdot\cos
2k_s z,
\end{equation}
where $n^{(1)}=N/L$. We suppose that $|\delta n|/n^{(1)}\ll 1$.
The phase $\varphi_k$ does not enter into this expression owing to
the accepted relation $\varphi_k=\varphi_{-k}$.

It should be emphasized that the static periodical structure
(\ref{eq:29}) is generated by the pair correlations between
phonons and has the amplitude proportional to the number of
quasiparticles in the stationary phonon state with the macroscopic population.

{\bf 2.} Let us make some quantitative estimates. For
definiteness, we consider the quasi 1D situation. As is known, in
this case the chemical potential is equal to $\mu=\omega_{\bot}s$
where $s=a^{(3)} n^{(1)}$, $a^{(3)}$ is the 3D scattering length (see
\cite{O}). The quasi 1D situation takes place under condition $s<
1$. Determining the stationary magnitude of the number of phonons,
we use expressions (13), (16) and (\ref{eq:22}) supposing $E_0\gg
\gamma, \xi_k$. As a result, we find $$N_{k}^s\approx
\frac{E_0}{|\bar{A}|}\approx
N\Big(\frac{\omega_{\bot}}{\omega_0}\Big)\eta
s;\;\;\;\;\;\eta=\frac{|\delta \omega_{\bot}|}{\omega_{\bot}}.$$
Using this value and definition of the chemical potential, we
obtain the following expression for the amplitude of the periodic
term in the longitudinal condensate density $$\frac{|\delta
n|}{n^{(1)}}\approx
\Big(\frac{\omega_{\bot}}{\omega_0}\Big)^2s^2\eta.$$ To estimate
the number of the phonon levels $N_0$ within the parametric
interval $2E_0$, we take into account that the distance between
the nearest levels is equal to $\delta\omega_{\|}=2\pi c_0/L$,
where $c_0=\sqrt{\mu/2m}$ is the averaged value of the
longitudinal sound velocity. Then
$$N_0=\frac{2E_0}{\delta\omega_{\|}}=\frac{\sqrt{2}}{\pi}\frac{\eta}{\sqrt{s}}\cdot\frac{\omega_0}{\omega_{\bot}}
\cdot\frac{L}{l_{\bot}},$$ where
$l_{\bot}=\sqrt{1/m\omega_{\bot}}$. Let us consider a set of
parameters acceptable for experiment
\[
\begin{split}
N=&10^4;\;\;\;L=2\cdot10^{-2}cm;\;\;\;\omega_{\bot}=10^4
sec^{-1};\\ &\frac{\omega_0}{\omega_{\bot}}= 0.2;\;\;\;\;\eta=
10^{-2}.
\end{split}
\]
For rubidium atomic gas we arrive at the following estimates $$
s=0.25;\;\;N_{k}^s\approx 10^{2};\;\;\frac{|\delta n|}{n^{(1)}}\approx
10^{-1};\;\;\;N_0\sim 1.$$ We see that all conditions assumed
before are satisfied.

Thus we can conclude that the state with the macroscopic number of phonons can be realized
for a realistic set of parameters.

\section {VII. Conclusion}

The analysis presented in the paper has revealed the nontrivial
picture of the temporal evolution of phonons created in a
Bose-condensed atomic gas as a result of the parametric resonance.
The phonon-phonon interaction is shown to play a decisive role
here. The interaction leads to the renormalization of phonon
levels which increases with growing the number of phonons. It
turns out that in all cases the evolution stops, when one of
phonon levels within the parametric interval approaches the edge
of this interval (phonon losses are compensated by the permanent
pumping). The macroscopic population of a single level is
characteristic for this event. At the same time the population of
all other levels is asymptotically going to zero. Actually, the
stationary level is filled by pairs of correlated phonons with the
opposite momenta $k_s,-k_s$ and the common phase. Thus the
self-consistent nonlinear evolution under consideration has
completed with the formation of the condensate composed of two
parts. One part is still in the initial state. The second part is
made up of the macroscopic number of phonons and gives rise to the
modulation in the atomic density.

It should be noted that, for the finite but sufficiently low
temperatures $T\lesssim\hbar\omega_0$, when the initial number of
phonons with $\omega_k\sim\omega_0$ is limited, the results
obtained remain valid.

This work is supported by the Russian Foundation for Basic
Research (Grant No.07-02-00067a) and by the Grant for Russian
Science Schools (No.NS-6869.2006.2).

\end{document}